\documentclass[12pt, a4paper]{article}
\usepackage[margin=2.5cm]{geometry}
\usepackage[utf8]{inputenc}
\usepackage{amsmath}
\usepackage{amssymb}
\usepackage{graphicx}
\usepackage{bbm}
\usepackage{hyperref}
\hypersetup{
    colorlinks = true,
    urlcolor   = blue,
    citecolor  = black,
}
\usepackage{newtxtext}
\usepackage{newtxmath}

\usepackage{bm}
\usepackage{listings}
\usepackage{xcolor}

\usepackage{natbib}

\usepackage{float}

\usepackage{authblk}  % Para gestionar múltiples autores y afiliaciones

\newcommand{\B}{\mathcal{B}}
\newcommand{\Tr}{\operatorname{Tr}}

\title{\textbf{Topological origin of flow distributions in disordered porous media}}

\author[1,2]{Jose Arnal\thanks{Corresponding author: \texttt{jose.arnal@idaea.csic.es}}}
\author[3]{Guillem Sole-Mari}
\author[1]{Tom\'{a}s Aquino}

\affil[1]{\small Spanish National Research Council (IDAEA -- CSIC), 08034 Barcelona, Spain}
\affil[2]{\small Departament de F\'{i}sica de la Mat\`{e}ria Condensada, Facultat de F\'{i}sica, Universitat de Barcelona, 08028 Barcelona, Spain}
\affil[3]{\small Department of Civil and Environmental Engineering, Universitat Polit\`{e}nica de Catalunya, 08034 Barcelona, Spain}

\date{} % Elimina la fecha automática del sistema

\begin{document}
%\linenumbers
\maketitle

\begin{abstract}
  We investigate steady Stokes flow through porous media composed of
  two-dimensional disordered arrays of circular obstacles. We develop a theory 
  for the statistics of flow rates based on a pore-network
  model that captures local flow correlations. We show that the flow rate
  distribution across the ensemble of pore bodies follows a Gamma distribution,
  and that the flow rate distribution through pore throats is fully determined
  in terms of it. Furthermore, this Gamma distribution can be directly linked
  to simple geometrical properties such as the coefficient of variation of pore
  throat widths, rendering the model parameterisable from minimal medium
  information. The resulting predictions agree closely with computational fluid
  dynamics simulations and show markedly better agreement than prior mean-field
  models that neglect local flow-rate correlations, clarifying how local
  splitting and merging shape flow in disordered porous networks.
\end{abstract}

\section{Introduction}

Flow and transport in porous media are fundamental to a wide variety of
processes and applications, including the characterisation and management of
groundwater resources and drinking water supplies \citep{UnitedNations2022,
  cherry1979groundwater, shannon2008}, oil and natural gas extraction
\citep{blunt2017, orr1984use}, $\mathrm{CO}_2$ sequestration
\citep{benson2008co2, szulczewski2012lifetime}, hydrogen storage
\citep{heinemann2021enabling}, industrial filtration processes
\citep{herzig1970flow}, chemical reactors \citep{froment1990chemical}, and
battery and fuel cell design \citep{manthiram2014, wang2004fundamental}.
However, the cross-scale heterogeneity that typically characterises both
natural and engineered porous media poses significant challenges to their study
\citep{sahimi2011flow, bear2013dynamics, blunt2017}. Even in Darcy-, or
continuum-, scale models, where local variability is spatially averaged, there
remain significant open challenges, such as quantitatively predicting transport
properties \textit{a priori} from medium characteristics, and deriving robust
upscaled models for reaction and mixing \citep{berkowitz2006modeling,
  dentz2011mixing, le2008lagrangian}. In microscopic, pore-scale models,
further challenges arise: the intricate complexity of the pore space introduces
a level of detail that renders full, point-by-point solutions of the flow field
intractable for domain sizes that are relevant at larger scales. Fortunately,
such detailed knowledge is typically not necessary: instead, a statistical
treatment of pore-scale velocities and flow rates is often sufficient to
capture essential transport phenomena. Indeed, the importance of point flow
statistics lies not only in the fact that they are an effective way to
characterise how flow is organised and partitioned within a porous medium, but
also because they can be used to inform stochastic transport and mixing models
based on continuous time random walks~\citep{berkowitz1998theory, dentz2004time}
or spatial-Markov processes~\citep{le2008lagrangian, dentz2016continuous, aquino2021diffusing, sherman2021review}, which have proven
very successful in predicting transport phenomena across scales. However,
predictive models linking the geometric and topological structure to flow statistics in heterogeneous porous media are still severely limited,
which in turn limits the predictive parameterisation of such stochastic models.
% In this work,
% we present a model that satisfactorily predicts flow rate distributions, both
% in pore bodies and throats, within a disordered two-dimensional collection of
% solid circular obstacles.

Many classical attempts to predict flow statistics in porous media involve
reducing the medium's heterogeneity to point statistics of the geometric
disorder, that is, to the variability in pore and throat sizes, while ignoring
how these pores are interconnected. An example of this approach is the
\textit{capillary bundle model}, in which the medium is represented as a
collection of parallel, independent tubes of different radii traversing the
medium. Despite its simplicity, this model has proven surprisingly useful in
determining certain properties of porous media. For instance, it has been shown
to be useful for calculating permeability and reconstructing the pore size
distribution from experimental capillary pressure curves
\citep{purcell1949capillary, fatt1956network, kozeny1927ueber, carman1937fluid},
as well as for predicting the probability density of very low flow velocities in
stagnation zones in media with broad pore size distributions
\citep{de2017prediction}. In the capillary bundle model, the flow rate
distribution in the tubes can be easily calculated from the
distribution of radii, which can sometimes be measured experimentally
\citep{dullien2012porous} and can often be well approximated by a unimodal probability density function (PDF),
such as a lognormal, Gaussian, or Gamma distribution
\citep{brutsaert1966probability, kosugi1994three}. This type of model is thus
easy to apply and often leads to analytical results with a clear physical
interpretation \citep{kutsovsky1996nmr, de2017prediction}. However, the
assumption that all tubes are mutually independent and that flow variability
depends solely on tube-width heterogeneity fails dramatically when attempting
to reconstruct flow statistics, even in relatively simple media \citep{alim2017local}. 

A more complex type of model that nonetheless retains some of the simplicity of
the capillary bundle model is the \textit{pore network model} (PNM), which
accounts not only for point statistics of geometry but also for the topology of
the pore space. This type of approach was first introduced by
\citet{fatt1956network}. In this framework, flow occurs through a collection of
nodes and links: the pore space is reduced to a graph where the sites represent
pore bodies and the bonds represent the throats connecting them. The pore
network model is capable of satisfactorily explaining and predicting a large
number of phenomena that are outside the scope of the capillary bundle model,
such as the emergence of preferential flow paths and anomalous dispersion in
single-phase transport \citep{kang2011spatial, dentz2018mechanisms, liu2024scaling, dentz2015mixing}, as well as hysteresis in oil extraction processes and
relative permeabilities in multiphase flow \citep{fatt1956network,
  valvatne2004predictive, golden1980percolation, reis2023simplified}. Perhaps one of its greatest
achievements, however, is that it allows for solving for flow with a
considerable level of precision at a computational cost far lower than that of
direct simulation methods: in PNMs, the computationally expensive task of
numerically integrating flow equations is replaced by the much less demanding
task of solving a linear system of equations (essentially, Kirchhoff's circuit
laws). 

While these laws are formulated in terms of total flow rates through pores and throats, which constitute the main focus of the present work, it is important to note that the stochastic
transport models mentioned above rely on velocity statistics, and that
intrapore velocity variability is known to play an important role in governing
transport phenomena such as dispersion~\citep{liu2024scaling, liu2026mechanism}. The translation
from flow rates to velocities is typically achieved by applying an appropriate
intrapore velocity stencil. A planar or cylindrical tube with a Poiseuille flow
profile is commonly employed, but more sophisticated models such as
doubly-parabolic throats can also be used to capture further microscopic
details \citep{ben2024pore}.

In addition to numerical solution of flow, one can ask whether the pore-network approach can yield general
predictions of flow statistics given incomplete knowledge of
the precise graph structure.
In the PNM representation, however, the analytical
calculation of the flow rate distribution is not as straightforward as in the
capillary bundle model. Indeed, even for simple geometries, the link between
structure and flow statistics remains poorly understood.
% One cannot arrive at an analytical expression
% as directly as by solving a simple single-variable integral, and needs to make
% some approximations.
% Despite the popularity
% This is the framework we will adopt: the one where the pore space is visualized
% as a graph.
In this context, the aim of the present work is to understand and quantify this
connection in simple disordered geometries. Namely, we focus on model porous
media composed of two-dimensional (2D) collections of circular solid obstacles,
where the fluid flows through the free spaces between these disks. This
idealised geometry is widely used across various disciplines in fluid mechanics
and transport phenomena, and it is classically referred to as a random array of
cylinders, a 2D disk pack, or a micropillar array (in experimental
microfluidics) \citep{sangani1988transport, cieplak1988dynamical,
  ishino2007wicking}. Despite its simplicity, and although some analytical
works exist regarding the completely ordered case \citep{hasimoto1959periodic,
  sangani1982slow}, there is no simple model that predicts how flow is
distributed and organised in the general, disordered case. Spatial disorder
breaks the symmetry of the medium, destroying simple periodic flow solutions in
which flow rates across parallel channels are always equal. Physically, this
leads to significantly higher tortuosity and broader flow rate distributions;
mathematically, it renders full analytical solutions intractable and makes a
statistical description of the flow necessary.

Previous
pore-network models for flow statistics in disordered systems employ what
we might call a \textit{mean-field approximation} \citep{coppersmith1996model,
  alim2017local}. It consists of assuming that all tubes in the network have
the same number of nearest upstream neighbours and that their flow rates are
statistically independent and identically distributed, with the splitting fractions determining the contribution of each upstream neighbour also being sampled independently from the same fraction distribution. As a consequence, Kirchhoff's first law (flow conservation at each junction) is only satisfied \textit{on average}, rather than being strictly enforced at every individual node. Based on this description, an analytical expression can
be derived for the distribution of the flow rate distribution in the tubes.
However, despite preserving some information about the network topology, in this work we
find that the mean-field approximation breaks down when predicting the throat
flow rate distribution for 2D arrays of circular obstacles with a high level of
spatial disorder.
As we will see, in two-dimensional arrays of disordered obstacles, some tubes
have a single upstream neighbour and others have two, and the statistical
properties of the flow rates for these two cases are fundamentally different.
Based on this distinction, we construct a model that is more accurate and
realistic than the mean-field approach, and which satisfactorily predicts flow
rate distributions in disordered two-dimensional porous media. We name this
model the \textit{Y} model, due to the fact that its fundamental building blocks consist
of three tubes joined at a node.

The model is laid out in detail in Section \ref{sec:theYmodel}. In Section
\ref{sec:parameters}, we derive a relationship between the model parameters and
the network geometry. In Section \ref{sec:validation}, we compare the model's
predictions with CFD and PNM simulations and with the predictions of the
mean-field model. Finally, Section \ref{sec:conclusion} provides a discussion
of the results and concludes the paper.

\section{The \textit{Y} model}
\label{sec:theYmodel}

As mentioned in the Introduction, we consider two-dimensional porous media
consisting of disordered collections of circular solid obstacles that do not touch or overlap.
% but are densely packed.
An incompressible fluid saturates the pore space between the disks and flows under
a steady-state Stokes regime. Macroscopically, the flow is directed along the
positive $x$ axis.

\begin{figure}
    \centering    \includegraphics[width=0.9\textwidth]{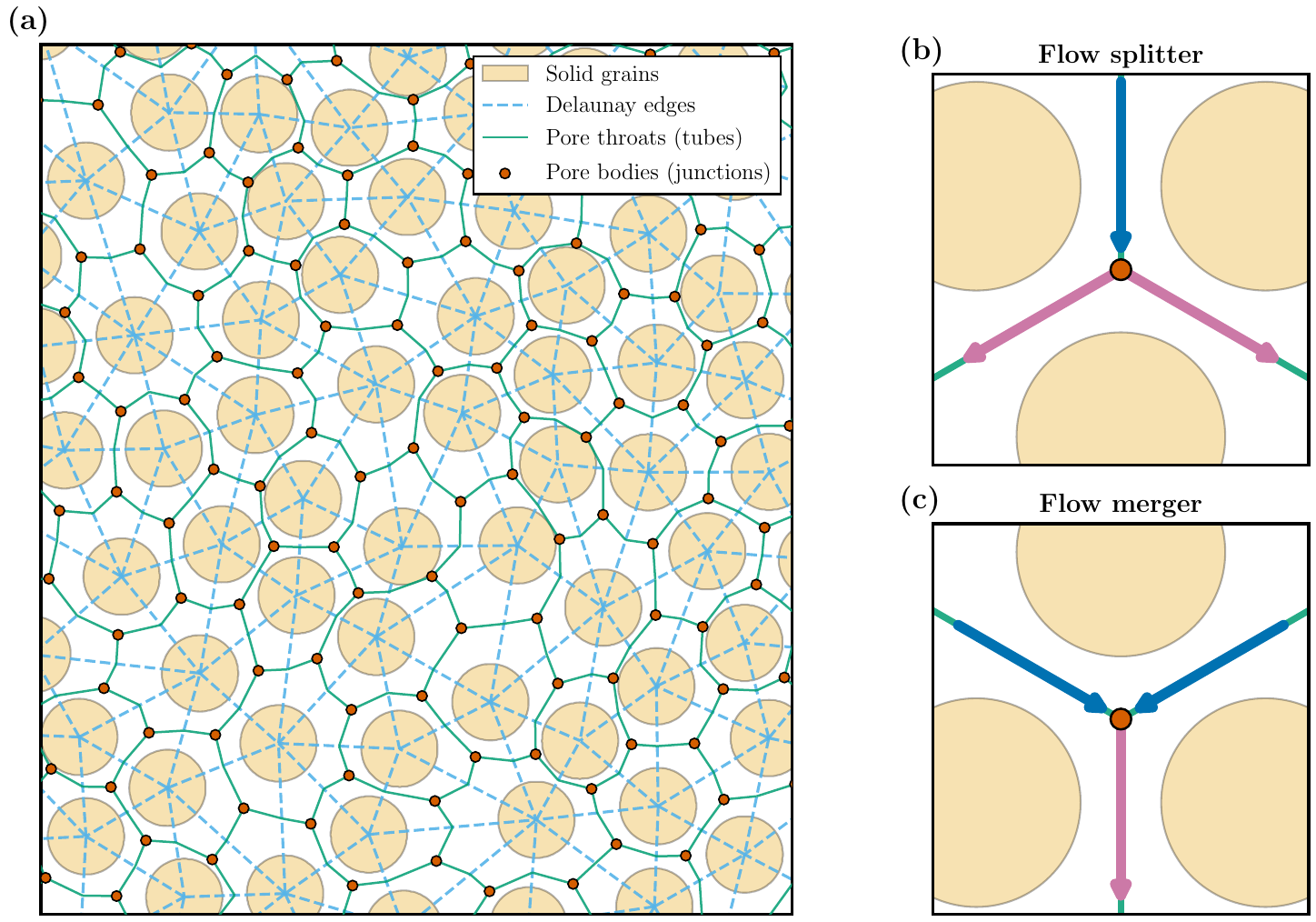}
    \caption{\textbf{(a)} Visualisation of the Delaunay tessellation and associated pore network. Each triangle corresponds to a pore body or junction. The geometric centroids of the triangles are used here to represent the junctions. While it might seem most natural to place these junctions at the circumcentres of the Delaunay triangles (as they coincide with the vertices of the Voronoi diagram), highly disordered media can cause the circumcentres to fall outside their respective triangles. This choice is purely illustrative: the important topological feature is that each triangle corresponds to a junction, and this junction conserves mass when exchanging flow with its adjacent neighbours. \textbf{(b, c)} Schematics of a flow splitter and a flow merger, respectively. These \textit{Y}-shaped structures represent the only two types of junction in the bulk of the network.}
    \label{fig:delaunay_and_network} 
\end{figure}

To identify the pore throats and pore bodies, we tessellate the porous medium
using a Delaunay triangulation: the plane is partitioned into non-overlapping
triangles whose vertices are the centres of the disks \citep{aurenhammer1991voronoi, kerstein1983equivalence}.
We identify the
narrow space between two neighbouring disks as a \textit{pore throat} or
\textit{tube}. The cross-section (or
cross-sectional segment, since the medium is 2D) of each throat is contained
within the edge of the Delaunay triangulation connecting the centres of its two
delimiting disks; specifically, it is the portion of that line segment lying
within the pore space. The throat flow rate is defined as the volume of fluid
going through its cross-sectional segment and, in 2D, has units of length
squared per time. We take flow rate values to be always positive by convention.
Each Delaunay triangle defines a distinct \textit{pore body} or \textit{junction} (see figure~\ref{fig:delaunay_and_network}). These correspond to points in the pore space where more than one
tube converges, or, in other words, to the \textit{sites} of the pore network. The specific geometric coordinate chosen to visually represent this junction is irrelevant, but we remark that standard choices have notable limitations: circumcentres may fall outside the triangle in highly disordered media, whereas geometric centroids may lie inside solid grains in polydisperse systems.

This definition of throats and pores based on Delaunay triangulation renders
the construction of the pore network straightforward. We define the flow rate of a pore body simply as the sum of the flow rates
of the tubes entering it, which, due to fluid incompressibility, is equal to
the sum of the flow rates of the tubes leaving it. Nodes associated with two
pores, $i$ and $j$, are connected by a link if the triangles associated with
$i$ and $j$ are adjacent. Furthermore, this link is identified with the throat
whose cross-sectional segment is contained within the Delaunay edge shared by
the two triangles. The direction of the link (whether from $i$ to $j$ or vice
versa) is determined by the direction of the flow within the throat.

As each pore or junction corresponds to a triangle, all junctions connect exactly three throats, resulting
in a regular pore network. To clarify that this is simply a counter-intuitive feature of the medium,
%and not a symptom of a flawed pore identification mechanism,
let us
examine the tessellation of the plane that is dual to the Delaunay
triangulation: the Voronoi diagram~\citep{aurenhammer1991voronoi}. The circumcentres of the triangles in a
Delaunay triangulation are the vertices of the Voronoi diagram for the same set
of generating points. Therefore, defining the pores as the vertices of the Voronoi diagram of the disk centres and the tubes as its edges would have yielded exactly the same network. The Voronoi
diagram partitions the plane into regions based on proximity to the disk
centres. Thus, the edges (corresponding to the tubes) consist of points
equidistant from two centres, while the vertices (corresponding to the pores)
correspond to points equidistant from three or more centres. However, three
points are sufficient to define a circle, and in 2D, the locus of points
equidistant from three points is the centre of the unique circle defined by
them. Consequently, for a Voronoi vertex to be equidistant from $n>3$
generating points (and thus connect $n>3$ edges or tubes) the centres of the
circles defined by any combination of three of them would have to coincide.
While this can easily be the case in regular lattices, in typical models of disordered media it happens with probability zero. In general, the same argument shows
that in disordered Voronoi networks in dimension $d$, the coordination number
is $z = d + 1$ \citep{jerauld1984percolation}. 

This fundamental topological property that all pores connect three throats is
one of the key ingredients of our model. Combined with local flow conservation, it implies that there exist
only two types of junctions between tubes, which we may call flow
\textit{mergers} and \textit{splitters} (see figure~\ref{fig:delaunay_and_network}). The \textit{Y} shape
characterising both types of junctions gives the model its name. 
In what follows, we
compute the flow rate statistics based on these junctions as the fundamental building block, as opposed to the single tube as in mean-field models~\citep{alim2017local}.

Having established that the pore networks under study are regular degree-3 graphs, and thus topologically homogeneous, a clarification regarding our use of the term \textit{disordered} seems necessary. While we may occasionally refer to geometric or topological variability as forms of disorder, when the term \textit{disordered} is used without qualification throughout this work, it refers specifically
to randomness in the obstacle centres, whose spatial arrangement strongly deviates from any regular lattice. It is important to distinguish this spatial randomness
from both topological and geometric heterogeneity. Topologically, the network is, as detailed above, highly regular, forming a degree-3 graph composed
strictly of simple \textit{Y}-shaped junctions. Geometrically, it typically exhibits a low level of size variability, with both obstacles and throat widths following relatively narrow distributions.

The flow rate through each pore and throat in the medium is not intrinsically
random: it is fully determined by the Stokes equations and the boundary
conditions. Nevertheless, the flow properties are highly sensitive to the
(quenched) geometric disorder. We thus conceptualise the flow rates in pores and
throats as random variables defined over the ensemble of all possible macroscopically equivalent realisations of this geometric disorder (i.e., independent configurations sharing the same global porosity and grain-size distribution). To assign a flow rate random variable
$Q_i$ to each specific pore $i$, we would need to track pore $i$
across different realisations of the disorder. However, since both the number
of pores and their positions vary significantly from one realisation to
another, this cannot be done. Instead, we adopt an alternative approach: we
define the pore ensemble as the infinite family consisting of all pores from
all possible disorder realisations, and we define the random variable $Q_p$ as
the flow rate of a pore chosen uniformly at random from this ensemble. This is a continuous random variable with positive support,
and we denote its PDF by $P_p(q)$. The subscript $p$ stands for \textit{pore},
and we refer to $P_p(q)$ as the \textit{pore flow rate distribution}. We define
the tube ensemble analogously, and we denote by $Q_t$ the random variable
that describes the flow rate of a tube chosen uniformly at random
%from this ensemble
and by $P_t(q)$ the corresponding \textit{throat flow rate distribution}.

Thus, the pore flow rates in a given network are realisations of $Q_p$, and the
tube flow rates are realisations of $Q_t$. For a specific realisation
of the geometric disorder, these realisations are not independent, as they are
strongly correlated locally due to flow conservation (Kirchhoff's first law).
However, we assume that these topological and spatial correlations decay at
large distances. Therefore, just as a single sufficiently large geometry
realisation allows us to sample the full variability of the disorder, it also
allows us to sample the full distribution $P_p(q)$, and likewise for $P_t(q)$.
In other words, we assume spatial ergodicity.

It should be noted that the standard notion of point statistics of flow rates coincides with the throat flow rate PDF as defined here. As is common in pore network models, the throat flow rate PDF describes the full flow statistics, without regard to the type of throat. As we will see in the remainder of this section, the pore flow rates may be seen as a formal device that allows us to make use of theoretical results that require distinguishing throats that have a splitting junction or a merging junction upstream.

\subsection{Pore flow rate distribution}
\label{sec:pore_flow_rate_distribution}

In this section, we derive an analytical expression for the pore flow rate
distribution $P_p(q)$. Let us consider an arbitrary pore $i$ with flow rate
$q$, representing a realisation of the random variable $Q_p$. Let us also
consider a downstream cross-section $S$ of the medium: a collection of pores
located within a transverse strip perpendicular to the macroscopic flow
direction that separates the medium's inlet from its outlet. Since
all paths connecting the inlet to the outlet must pass through at
least one node in $S$, the flow passing through
pore $i$ must traverse one of the pores in $S$ before reaching the outlet. Let
$V_i \subset S$ denote the subset of pores in the cross-section that receive a
non-zero fraction of the flow originating from $i$, and let $t_j$ denote the
partial flow rate that pore $j \in V_i$ receives from $i$. By flow
conservation, it follows that
%
%\begin{equation}
$q = \sum_{j \in V_i} t_j$.
%\end{equation}

We treat the partial flow 
rates $t_j$ as realisations of random variables $T_j$, analogously to how $q$
is a realisation of $Q_p$. In a disordered medium, it is reasonable to assume that if the cross-section
$S$ is located sufficiently far downstream from pore $i$, the partial flow
rates $T_j$ become statistically independent, in the sense that the value of
any specific value of $T_j$ provides no information regarding the values of the others,
other than the constraint that they must sum to $q$. Let us now consider the fraction $W_j$ that each
partial flow rate $T_j$ contributes to the total flow of pore $i$, defined as
%
%\begin{equation}
% W_j = \frac{T_j}{Q_p},
%\end{equation}
$W_j = T_j/Q_p$,
with $\sum_j W_j = 1$. Since Stokes flow is linear, we expect these fractions
to be independent of the total flow rate $Q_p$. In other words, redistribution of flow should depend solely on geometric and
topological properties, and not on the magnitude of the flow itself: if the
medium partitions a flow $q$ into portions $t_1, \dots, t_m$, it should
partition a flow $\lambda q$, $\lambda > 0$, into portions
$\lambda t_1, \dots, \lambda t_m$.

Lukacs's proportion-sum independence theorem, see \citet{lukacs1955characterization} and Appendix~\ref{app:lukacs_theorem}, states that, assuming the partial flow rates
$T_j$ are mutually independent, the splitting fractions $W_j$ are independent
of the sum $Q_p = \Sigma_j T_j$ if and only if the probability distribution of
$Q_p$ is a Gamma distribution,
\begin{equation}
\label{eq:Pp}
    P_p(q) = \frac{q^{k-1} e^{-q/\theta}}{\Gamma(k) \theta^k}.
\end{equation}
% Direct numerical simulations confirm this distribution for media exhibiting
% sufficient disorder.
The Gamma distribution is characterised by two parameters: the shape parameter
$k$ (which is dimensionless) and the scale parameter $\theta$ (which, in this
context, has units of 2D flow rate). Lukacs's theorem does not specify the
values of these parameters. We show in Section~\ref{sec:parameters} how
they can be computed in terms of geometric properties of the medium. 

\subsection{Throat flow rate distribution}
\label{sec:tube_flow_rate_distribution}

The \textit{Y} shape of the junctions occurring at the pores establishes a
direct relationship between the distributions $P_p(q)$ and $P_t(q)$, as we now
show.
Each tube or throat is bounded by two pores, carrying fluid from one
(which we call the \textit{inlet}) to the other (the \textit{outlet}).
Throughout this work, the terms \textit{inlet} and \textit{outlet} are used to
designate both the pores at either end of a single tube and the macroscopic
boundary pores of the system. While the distinction is typically evident from context, we explicitly specify \textit{tube inlet} or \textit{global inlet} (and likewise for outlets) whenever ambiguity might arise.

Let $\beta$ be the fraction of tubes whose inlet node is a merger (see figure~\ref{fig:delaunay_and_network}). The flow rate of such tubes is equal to that of their inlet node, which in
turn is a realisation of the random variable $Q_p$ introduced above. The other
possibility is that the inlet node of a tube is a splitter, which occurs for a
fraction $1 - \beta$ of the tubes. In that case, the tube's flow rate is a
fraction of its inlet node's flow rate. We represent this fraction by the random variable $\Omega$
defined over the ensemble of all splitter pores from all possible geometric
disorder realisations. The random variable $Q_t$ is thus a mixture of two
random variables:
\begin{equation}
Q_t = %
\begin{cases} Q_p, & \text{with probability } \beta, \\ \Omega   Q_p, & \text{with probability } 1 - \beta. \end{cases}
\end{equation}
Thus, if we denote by $P_s(q)$ the PDF of the random variable
$Q_s = \Omega   Q_p$ (where the subscript $s$ stands for \textit{split}), we
see that:
\begin{equation}
  P_t(q) = \beta  P_p(q) + (1-\beta)  P_s(q).
\end{equation}
A simple reasoning (see
Appendix \ref{app:probabilities}) shows that, as long as the number of global
inlet and outlet nodes is negligible, the number of merger pores must be
approximately equal to the number of splitter nodes. When this holds, the
probability $\beta$ that the inlet node of a uniformly chosen tube is a merger
is $1/3$.

We now derive an analytical expression for the PDF $P_s(q)$ of the flow rates
$Q_s=\Omega Q_p$ associated with tubes exiting splitter pores. First, note that
the pore flow rates $Q_p$ follow the Gamma distribution described in
Section~\ref{sec:pore_flow_rate_distribution}. Then, direct numerical simulations in sufficiently disordered disk packings
(see Section~\ref{sec:validation}) show that, while splitting fractions in nearby
tubes are not independent, the point
statistics of $\Omega$ are well approximated by those of a uniform random
variable $\mathcal{U}(0,1)$ with support in $[0,1]$. Using the standard formula
for the distribution of the product of two independent random variables, we
obtain
\begin{equation}
  P_s(q) = \int_0^\infty\frac{P_p(q)}{q'} P_\Omega(q/q')  \,dq'=\int_q^\infty \frac{P_p(q)}{q'}  \,dq' = \frac{\Gamma(k-1,   q/\theta)}{\Gamma(k) \theta},
  \label{eq:pdf_of_product}
\end{equation}
where $\Gamma(k-1,   q/\theta)$ is the upper incomplete Gamma function.
Thus, the PDF of the tube flow rates $Q_t$ finally takes the form
\begin{equation}
\label{eq:Pt}
    P_t(q) = \frac{1}{\Gamma(k) \theta} \left[ \frac{1}{3} \left(\frac{q}{\theta}\right)^{k-1} e^{-q/\theta} + \frac{2}{3} \Gamma\left(k-1,  \frac{q}{\theta}\right)  \right].
\end{equation}

The derivation of Equation (\ref{eq:Pt}) relies on the assumption that the splitting fraction $\Omega$ is uniformly distributed in $[0,1]$. This holds remarkably well for media with a reasonable degree of spatial disorder. Even in some strongly constrained geometries (such as the $\varepsilon$-perturbed medium with $\varepsilon = 0.9$ introduced in Section~\ref{sec:validation}) where the spatial arrangement of the obstacle centres departs only slightly from a regular lattice, the uniform distribution of splitting fractions is still observed. In more highly ordered geometries, however, the assumption of uniformly distributed splitting fractions breaks down. Nevertheless, the Gamma distribution for pore flow rates remains valid, and an analytical expression for the tube PDF can still be obtained by fitting the splitting fractions to a symmetric Beta distribution using the method of moments (see Appendix~\ref{app:ordered_limit}).

\section{Connecting distribution parameters to medium geometry}
\label{sec:parameters}

The pore and tube flow rate distributions, Eqs.~\eqref{eq:Pp}
and~\eqref{eq:Pt}, are both fully determined in terms of the Gamma
distribution's parameters $k$ and $\theta$. These are connected to the mean
$\langle \cdot \rangle$ and coefficient of variation
$\operatorname{CV}(\cdot) = \sqrt{\operatorname{Var}(\cdot)}/\langle\cdot\rangle = \sqrt{\langle\cdot^2\rangle -
  \langle\cdot\rangle^2}/\langle\cdot\rangle$ of the pore flow rate $Q_p$ by
\begin{equation}
    k = \frac{\langle Q_p \rangle ^2}{\operatorname{Var}(Q_p)} = \frac{1}{\operatorname{CV}^2(Q_p)},
\label{eq:k_in_terms_of_statistics}
\end{equation}
\begin{equation}
    \theta = \frac{\operatorname{Var}(Q_p)}{\langle Q_p \rangle} = \langle Q_p \rangle     \operatorname{CV}^2(Q_p).
\label{eq:theta_in_terms_of_statistics}
\end{equation}
In what follows, we close the model by connecting these quantities to geometric
properties of the medium.

\subsection{Mean pore flow rate}
\label{sec:parameters:mean}

The mean flow rate $\langle Q_p\rangle$ in the pores can be connected to the
total flow rate $q_{\mathrm{tot}}$ entering a disordered medium composed of
circular objects by simple geometric arguments. First, we estimate the number
of disks, $N_d$, within the domain. Since the disks do not overlap, we 
estimate the total surface area they occupy as $N_d$ times the average area $\pi \langle R^2 \rangle$ of
a disk, where $R$ is a random variable
describing the disk radii. This total solid area must be equal to $(1-\phi)\mathcal{V}$,
where $\phi$ is the porosity and $\mathcal{V}$ is the domain area (assumed much larger
than the disk area).

This yields
%
% \begin{equation}
%   N_d = \frac{(1-\phi)V}{\pi \langle R^2 \rangle}.
%\end{equation}
$N_d = (1-\phi)\mathcal{V}/(\pi \langle R^2\rangle)$.
Therefore, there are $\rho_d=(1-\phi)/\pi\langle R^2\rangle$ disks per unit area, and we can estimate their linear density as $\sqrt{\rho_d}$. At this point, we consider two limiting idealised scenarios to estimate the number of pores $N_w$ that lie immediately downstream of any arbitrary cross-section of the medium, which will provide bounds for the mean flow rate.

For our first estimate, we approximate that a cross-section of the medium of width $\mathcal W$ 
crosses, on average, a number of disks equal to $\mathcal{W}\sqrt{\rho_d}$. Between any two adjacent disks along this transverse line, there is a gap acting as a pore throat, resulting in approximately $\mathcal{W}\sqrt{\rho_d}$ such throats crossing the section. If we make the simplifying assumption that most of these throats lead to distinct pores, we conclude that the 
number of pores just after this section is also 
%\begin{equation}
%    N_w = \mathcal{W} \sqrt{\frac{1-\phi}{\pi \langle R^2 \rangle}}.
%\end{equation}
$N_w = \mathcal{W}\sqrt{\rho_d}=\mathcal{W}  \sqrt{(1-\phi)/(\pi \langle R^2 \rangle)}$.
Due to conservation of flow, the sum of the flow rates through these pores must
equal the total flow rate through the medium, $q_{\textrm{tot}}$. Hence, we can
estimate the mean pore flow rate as
\begin{equation}
\langle Q_p \rangle = \frac{q_{\textrm{tot}}}{N_w}
  = \frac{q_{\textrm{in}}}{\mathcal{W}} \sqrt{\frac{\pi \langle R^2 \rangle}{1-\phi}} =:
q_p^\ell,
  \label{eq:lower_bound}
\end{equation}
where $\ell$ stands for \textit{lower} because, as we will see shortly, this provides a lower bound for the mean flow rate.
Despite its simplicity, this geometric idealisation predicts the mean flow rate at the pores remarkably well. Validation against direct numerical simulations, see Section~\ref{sec:validation}, yields relative errors between 5\% and 13\%. 

For our second estimate, we perform a more refined computation of the number of distinct pores $N_w$ immediately downstream of the cross-section. We start with the same number of intersecting throats, $\mathcal{W}\sqrt{\rho_d}$, but we now assume that the set of their endpoints follows the general topological statistics of the network. As discussed in Appendix~\ref{app:probabilities}, in a regular degree-3 pore network, half of the nodes act as mergers (two incoming tubes, one outgoing tube, see figure~\ref{fig:delaunay_and_network}) and half act as splitters (one incoming tube, two outgoing tubes).
Consequently, $2/3$ of the crossed tubes will share their downstream endpoint in pairs, while the remaining $1/3$ will connect to distinct individual pores. From this, we can compute the number of distinct pores $N_w$ lying immediately downstream of the section:
\begin{equation}
    N_w = \frac{1}{2}\left(\frac{2}{3}\mathcal{W}\sqrt{\rho_d}\right) + 1\left(\frac{1}{3}\mathcal{W}\sqrt{\rho_d}\right) = \frac{2}{3}\mathcal{W} \sqrt{\frac{1-\phi}{\pi \langle R^2 \rangle}},
\end{equation}
where we have estimated again the number of tubes in the cross-section as $\mathcal{W}\sqrt{\rho_d}$.
Proceeding with the conservation of mass as before, this reduced number of effective pores leads to our second estimate:
\begin{equation}
    \langle Q_p \rangle = \frac{3}{2} \frac{q_{\textrm{tot}}}{\mathcal{W}} \sqrt{\frac{\pi \langle R^2 \rangle}{1-\phi}} =: q_p^{u} = \frac{3}{2} q^\ell_p,
    \label{eq:upper_bound}
\end{equation}
where $u$ stands for \textit{upper}.

DNS results show that these two estimates bracket the true mean flow rate at the pores (see Section~\ref{sec:validation}). The assumption that every throat leads to a distinct pore slightly overestimates the number of pores immediately downstream of the cross-section, yielding a lower bound $q_p^\ell$ for the mean flow. Conversely, assuming the global splitting and merging statistics described in Appendix~\ref{app:probabilities} are preserved in the cross section slightly underestimates the effective number of cross-section pores, yielding an upper bound $q_p^u$. The true average flow rate is always observed to fall between these two idealisations, which can be estimated solely from macroscopic parameters: the total flow rate $q_{\mathrm{tot}}$, the domain width $\mathcal{W}$, the porosity $\phi$, and the second moment of the disk radii.

\subsection{Coefficient of variation of the pore flow rate}
\label{sec:determination_of_CV}

Next, we relate the coefficient of variation of the pore flow rate distribution
to the coefficient of variation of the heterogeneity in throat widths. Here, we
employ a linearisation valid under the assumption of mild variability in throat
widths, as detailed below. We find in Section~\ref{sec:validation} that this
linearisation performs well for a wide range of scenarios. 

Once both the medium topology and the total flow are fixed, the values of the
flow rates in the pores only depend on the values of the tube conductances.
Assuming, for example, the Hagen-Poiseuille law for each tube, for a fixed
value of a tube's length, the conductance is itself completely determined by
the tube's half-width. Thus, for a specific network, the vector of pore flow
rates $\bm{q}_p\in\mathbb{R}^{N_p}$ is a function of the vector of half-widths
$\bm{a}\in\mathbb{R}^{N_t}$.

Let us consider an initially uniform geometry where all tubes share the same
half-width, $\bm{a} = \bm{a}_0 = (a_0, \ldots, a_0)$. We introduce geometric
disorder by applying a small, random perturbation $\delta \bm{a} = \bm{a} - \bm{a}_0$ to this
homogeneous state. To preserve the mean half-width, each local perturbation
$\delta a_j$ is sampled independently from a probability distribution with zero
mean. Consequently, the new half-widths, given by $a_j = a_0 + \delta a_j$, are
independent realisations of identically distributed random variables $A_j$ with
an expected value $\langle A_j\rangle = \left< A \right>= a_0$, where for
notation convenience we introduce a random variable $A$ with the same distribution. For
a given network topology, we distinguish between two types of average: first,
we retain angle brackets ($\langle \cdot \rangle$) in this context to represent
the ensemble average with respect to geometric (i.e., throat-width) variability only; and second, for a given realisation of the geometric
variability, we
employ an overline ($\overline{\cdot}$) to denote a spatial average over all
network elements.

We now evaluate the effect of geometric variability on the local flow
dynamics. In the linear limit (i.e., for infinitesimal perturbations
$\delta \bm{a}$), the new pore flow rate vector can be approximated by a
first-order Taylor expansion around the uniform state:

\begin{equation}
    \bm{q}_p(\bm{a}_0 + \delta \bm{a}) = \bm{q}_p^0 + J^0 \delta \bm{a},
    \label{eq:slight_heterogeneity}
\end{equation}
where $\bm{q}_p^0 = \bm{q}_p(\bm{a}_0)$ is the set of pore flow rates in the homogeneous case represented by $\bm{a}_0$, and $J^0$ is the Jacobian matrix of the vector function
$\bm{q}_p(\bm a)$ evaluated at $\bm{a}_0$,
\begin{equation}
    J^0_{ij} = \frac{\partial q_{p,i}}{\partial a_j}\bigg|_{\bm{a}=\bm{a}_0}.
\end{equation}
% The new half-widths $a_j$ are independently sampled from a certain
% distribution whose mean is $\bar a$.
% Then, the flow rate at node $i$ in the heterogeneous (i.e., perturbed) case is:
% % 
% \begin{equation}
%   q_{p,i} = q_{p,i}^0 + \sum_{j=1}^{N_t}\frac{\partial q_{p,i}}{\partial a_j}\bigg|_{\mathbf{a}=\mathbf{a}_0}(a_j-a_0).
%   \label{eq:slight_heterogeneity}
% \end{equation}

Since the half-widths $a_j$ are realisations of the random variables $A_j$, Eq.~\eqref{eq:slight_heterogeneity} also defines the random flow rates $Q_{p,i}$ simply by replacing $\delta \bm a$ by $\delta \bm A = \bm{A} - \bm{a}_0$. If we compute the first moment of both sides of the resulting equation with respect to geometric variability we find
\begin{equation}
    \langle Q_{p,i} \rangle = q_{p,i}^0 + \sum_{j=1}^{N_t}\frac{\partial q_{p,i}}{\partial a_j}\bigg|_{\mathbf{a}=\mathbf{a}_0}(\langle A_j \rangle-a_0) = q_{p,i}^0.
    \label{eq:mean_flow_tube}
\end{equation}
This means that a small perturbation that leaves the average throat width
invariant also leaves the mean pore flow rate unchanged. For the second moment we obtain
\begin{equation}
%
% \begin{split}
  \langle Q_{p,i}^2\rangle
  % &= \left\langle \left[q_{p,i}^0 + \sum_{j=1}^{N_t} J^0_{ij}(A_j-a_0)\right]^2\right\rangle, \\
  % &= \left\langle (q_{p,i}^0)^2 + \sum_{j=1}^{N_t}\sum_{k=1}^{N_t} J^0_{ij}J^0_{ik}(A_j-a_0)(A_k-a_0) + 2 q_{p,i}^0\sum_{j=1}^{N_t}J_{ij}^0(A_j-a_0)\right\rangle,\\
  % &=(q_{p,i}^0)^2 + \sum_{j=1}^{N_t}(J_{ij}^0)^2\langle(A-a_0)^2\rangle
  = (q_{p,i}^0)^2 + \operatorname{Var}(A)\sum_{j=1}^{N_t}(J_{ij}^0)^2,
  \label{eq:second_moment_flow_tube}
% \end{split}
\end{equation}
where the final expression follows from expanding the square, noting that $\langle A_j - a_0 \rangle = 0$, and using $\langle (A_j-a_0)(A_k-a_0)\rangle = \delta_{jk}\operatorname{Var}(A)$ under the assumption of spatially uncorrelated half-width perturbations.

To relate the variance of pore flow rates to the tube half-width variance, we must compute the first and second pore flow moments by averaging not only over the distribution of $A$ for a given topology ($\langle \cdot \rangle$), but also over all possible topology realisations. Under the assumption of spatial ergodicity mentioned in Section~\ref{sec:theYmodel}, this topological ensemble average can be closely approximated by a spatial average over all elements within a single network realisation ($\overline{\cdot}$). Thus, using~\eqref{eq:mean_flow_tube} and~\eqref{eq:second_moment_flow_tube},

\begin{equation}
\begin{split}
    \operatorname{Var}(Q_p)
    &
    = \langle \overline{Q_p^2} \rangle- \langle \overline{Q_p} \rangle^2
    =\frac{1}{N_p}\sum_{i=1}^{N_p}\langle Q_{p,i}^2 \rangle  -\left(\frac{1}{N_p}\sum_{i=1}^{N_p}\langle Q_{p,i}\rangle\right)^2\\
    % &= \frac{1}{N_p}\sum_i \left[(q_{p,i}^0)^2 + \frac{\operatorname{Var}(A)}{N_p}\sum_j(J_{ij}^0)^2\right]-\left(\frac{1}{N_p}\sum_{i=1}^{N_p} q_{p,i}^0\right)^2 \\
    &
    = \operatorname{Var}(Q_p^0) + \frac{\operatorname{Var}(A)}{N_p} \|J^0\|_{\textrm{F}}^2,
\end{split}
\end{equation}
where $Q_p^0$ is a random variable describing the flow rates at the pores in
the homogeneous case, and $\|M\|_\textrm{F}=\sqrt{\Tr(MM^T)} = \sum_{i,j=0}^{N_t}|M_{ij}|^2$ is the so-called
\textit{Frobenius norm} of a matrix $M$, with $\Tr(\cdot)$ denoting the trace operator and the superscript $T$ indicating transposition.
% (the Frobenius norm of a matrix is simply the square root of the sum of the
% squares of the moduli of its entries. Here we have the Frobenius norm of $J$
% squared).
Dividing by $\overline{q_p^0}^2$, with $\overline{q_p^0} = \sum_iq^0_{p,i}/N_p$,
we obtain
\begin{equation}
  \operatorname{CV}^2(Q_p) = \operatorname{CV}^2(Q_p^0) +\frac{\|J^0\|_\textrm{F}^2 
    \left< A \right>^2}{N_p \overline{q_p^0}^2} \operatorname{CV}^2(A).
  \label{eq:variances_relation}
\end{equation}
This is a powerful result in that it isolates the two sources of variability in
pore flow rates. On the one hand, we have what we might term
\textit{topological disorder}, represented by $\operatorname{CV}^2(Q_p^0)$,
which is present even when all tubes have the same conductance. On the other
hand, the remaining term encodes \textit{geometric disorder}, due solely to the
variability in tube half-widths.

Eq.~\eqref{eq:variances_relation} is a first-order approximation in
half-width perturbations that is exact only in the infinitesimal limit. This
means that it holds only when the geometric disorder is
sufficiently mild (more specifically, when the dimensionless quantity
$\operatorname{CV}^2(A)$ is sufficiently small). This is, in practice, usually
the case for densely-packed arrays of disks, see Section~\ref{sec:validation}. In addition,
the numerical simulations in Section~\ref{sec:validation} indicate that the topological disorder contribution represented by
$\operatorname{CV}^2(Q_p^0)$ is very small and can be neglected, so that $\operatorname{CV}^2(Q_p)$ turns out to be dominated by the geometric term even for small
half-width variability.
%Thus, $\operatorname{CV}^2(Q_p)$ turns out to be dominated by the geometric term.
Apart from these considerations, Eq.~\eqref{eq:variances_relation} is not
limited to this type of medium.

%We now use some properties 2D disordered arrays of disks to
% particularize the result for this scenario. \comTA{It seems to me that what follows is still general?}
We thus turn to the geometric contribution.
%, which turns out to be dominant for disordered disk packings,
% This term depends on the Frobenius norm of
% the Jacobian of the function that relates pore flow rates to tube widths.
We now show that, under generic assumptions on the disorder, the prefactor of $\operatorname{CV}^2(A)$ in Eq.~\eqref{eq:variances_relation} has a universal
numerical value that can be computed analytically. This means in particular that the geometric contribution to the flow rate variability indeed depends only on geometric (throat-width) variability.
To do so, let
$\bm{p}\in\mathbb{R}^{N_p}$ and $\bm{q}_t\in\mathbb{R}^{N_t}$ be column vectors
representing the pressure values at the nodes and the flow rates at the tubes.
Let also $\bm{s}\in\mathbb{R}^{N_p}$ be a column vector whose components are
the external flow rates, that is:
\begin{equation}
  s_i =
  \begin{cases}
    q_{\textrm{in}, i}, & \text{if $i$ is a global inlet,} \\
    -q_{\textrm{out}, i}, & \text{if $i$ is a global outlet,} \\
    0 & \text{otherwise},
  \end{cases}
  \label{eq:source_term}
\end{equation}
with the condition that $\sum_{i=1}^{N_p} s_i = 0$. We assume for simplicity
that the values $q_{\textrm{in}, i}$ and $q_{\textrm{out}, i}$ are directly imposed
as boundary conditions. In Eq.~\eqref{eq:source_term}, we
have adopted the convention that flow rates \textit{entering} a node are
\textit{positive}, whilst flow rates \textit{leaving} it are \textit{negative}.
This sign convention is arbitrary.

Next, let $C\in\mathcal{M}_{N_t\times N_t}(\mathbb{R})$ be a diagonal matrix where
$C_{jj}$ is the conductance of the $j$th tube, and let
$B\in\mathcal{M}_{N_p\times N_t}(\{0,1\})$ be the so-called \textit{incidence matrix}:
\begin{equation}
  B_{ij} =
  \begin{cases}
    -1, & \text{if tube $j$ leaves node $i$ ($i$ is the inlet node of tube $j$),} \\
    1, & \text{if tube $j$ enters node $i$ ($i$ is the outlet node of tube $j$),} \\
    0, & \text{otherwise},
  \end{cases}
\end{equation}
The elements of any column of $B$ always sum up to $0$, as every tube always
has an inlet node and an outlet node. On the other hand, the elements of the
$i$th row of $B$ sum up to $1$ if pore $i$ is a global inlet (no incoming tubes, one
outgoing tube) or a splitter (one incoming tube, two outgoing tubes), and to
$-1$ if it is a global outlet (no outgoing tubes, one
incoming tube) or a merger (one outgoing tube, two incoming tubes), see figure~\ref{fig:delaunay_and_network}.
% This is another way of proving that,
% whenever the number of inlet nodes is approximately equal to the number of
% outlet nodes, or whenever both numbers are negligible, the number of splitters

In terms of these quantities, the stationary state of the network is governed
by two linear equations, namely
\begin{subequations}
\begin{equation}
    B \bm{q}_t + \bm{s} = \bm{0}
    \label{eq:KCL}
\end{equation}
and
\begin{equation}
    \mu\bm{q}_t = -CB^T\bm{p},
    \label{eq:Ohm}
\end{equation}
\label{eq:KCL_Ohm}
\end{subequations}
where $\mu$ is the dynamic viscosity of the fluid. 
Eq.~\eqref{eq:KCL} is simply Kirchhoff's first law
for a hydraulic circuit, or flow conservation at each node.
% To see this more
% clearly, consider a (merger) bulk node $i$ that receives flow from tubes $j_1$
% and $j_2$, and delivers flow to tube $j_3$. For this node, Eq.~\eqref{eq:KCL}
% reads 
% %
% \begin{equation}
%      q_{t,j_1} +q_{t,j_2} -q_{t,j_3} = 0.
% \end{equation}
% Alternatively, if $i$ is an inlet node (so it receives flow only from outside
% the network and gives all its flow to tube $j$), Eq.~\eqref{eq:KCL} reads
% %
% \begin{equation}
%     -q_{t,j}+q_{\textrm{in},i}  = 0,
% \end{equation}
% and similarly for outlet nodes.
Eq.~\eqref{eq:Ohm} is the constitutive relation
of the system: it is the law that relates the flow rate at each tube to the
pressure drop through it via its conductance. Note that if $i_1$ is the inlet
pore of tube $j$ and $i_2$ is its outlet pore, the pressure drop (in absolute
value) through $j$ is $| p_{i_2}-p_{i_1}| = p_{i_1} -p_{i_2} $, which
coincides with the $j$th entry of the vector $-B^T\bm{p}$. Thus,
\begin{equation}
  \label{eq:pressure_drops}
  \Delta \bm{p} = -B^T\bm{p}
\end{equation}
is the vector of pressure drops, and consequently Eq.~\eqref{eq:Ohm} is the
statement of Darcy's law for each tube in matrix notation. Note that the conductance $C_{jj}$ of tube $j$ (which we often term \textit{geometric conductance} as it is fluid-independent) is directly related to the equivalent intrinsic permeability $k_j$ through the relation $C_{jj}=2a_jk_j/\ell_j$, where $\ell_j$ is the tube's length, and the standard hydraulic conductance is given by $C_{jj}/\mu$.

By substituting Eq.~\eqref{eq:Ohm} into~\eqref{eq:KCL}, we can solve for
$\bm{p}$ in terms of $\bm{s}$:
\begin{equation}
    \bm{p} = \mu\left[BCB^T\right]^{-1}\bm{s}.
    \label{eq:pressures_in_terms_of_sources}
\end{equation}
Note that because pressures are only defined up to an additive constant, $BCB^T$ is singular; we render it invertible by fixing a reference node (removing its row and column), slightly abusing notation by keeping the original matrix symbols for the reduced system.
Inserting Eq.~\eqref{eq:pressures_in_terms_of_sources} back into Eq.~\eqref{eq:Ohm} leads to
\begin{equation}
 \bm{q}_t = -CB^T\left[BCB^T\right]^{-1}\bm{s}.
 \label{eq:tube_flows_in_terms_of_sources}
\end{equation}
Interestingly, this result is independent of the dynamic viscosity $\mu$: given the flow input and output at the domain boundaries ($\bm{s}$), the flow is fully determined by the network topology ($B$) and geometry ($C$).

Eq.~\eqref{eq:variances_relation} involves the pore flow rates rather than the
throat flow rates. Due to the \textit{Y} shape of junctions, there is always a
tube either entering or leaving a pore that carries all its flow rate, and this remains true at global inlets and outlets.
% For
% global inlets and mergers, this tube is the single outgoing tube; for global
% outlets and splitters, it is the single incoming tube.
We can therefore define
an \textit{inclusion matrix} $Z\in\mathcal{M}_{N_p\times N_t}(\{0,1\})$ such
that the vector $\bm{q}_p\in\mathbb{R}^{N_p}$ of total flow rates through each
pore is given by
\begin{equation}
    \bm{q}_p = Z \bm{q}_t.
\end{equation}
Without loss of generality, network nodes can be labelled so that pore $i$
shares the same index as the single tube carrying its entire flow rate. Under
this numbering scheme, $Z$ is a rectangular $N_p\times N_t$ matrix with ones on
its main diagonal and zeros elsewhere. Irrespective of the numbering scheme, we have
\begin{equation}
    \bm{q}_p = -ZCB^T\left[BCB^T\right]^{-1}\bm{s}.
    \label{eq:pore_flows_in_terms_of_sources}
\end{equation}
This encodes the dependence of $\bm{q}_p$ on the half-widths $\bm{a}$ through the conductances $C$ (the geometric conductance of each tube is related to its half-width via, for example, the Hagen-Poiseuille law).

We are now in a position to compute the Jacobian $J^0$ in Eq.~\eqref{eq:variances_relation}, which represents the sensitivity of pore flow rates to changes in tube half-widths around the homogeneous network ($\bm{a} = \bm{a}_0$). By differentiating Eq.~\eqref{eq:pore_flows_in_terms_of_sources} with respect to the tube half-widths, we can obtain an exact analytical expression for each column of the Jacobian. As detailed in Appendix~\ref{app:computation_of_jacobian}, this calculation yields:

\begin{equation}
  \frac{\partial\bm{q}_p}{\partial a_j}\bigg|_{\bm{a}={\bm{a}_0}} = \frac{3q^0_{t,j}}{a_0} Z\left[\mathbbm{1}_t-H\right]\bm{e}_j,
  \label{eq:jacobian_column_homogeneous}
\end{equation}
with $H = B^T\left[BB^T\right]^{-1}B$, where $\mathbbm{1}_t$ is the $N_t \times N_t$ identity matrix and $\bm{e}_j$ is the $j$th vector of the standard basis in $\mathbb{R}^{N_t}$. Note that $H$ is symmetric ($H^T = H$) and idempotent ($H^2 = H$). In other words, $H$ is an orthogonal projector, and therefore so is $\mathbbm{1}_t - H$. This gives Eq.~\eqref{eq:jacobian_column_homogeneous} an elegant physical interpretation. From Eq.~\eqref{eq:KCL}, it follows that any flow perturbation $\delta\bm{q}_t$ resulting from network changes must satisfy the homogeneous Kirchhoff's first law $B \delta\bm{q}_t = \bm{0}$, to ensure that the new flow field continues to respect local flow conservation at each junction. Consequently, the null space of the incidence matrix $B$ represents the linear subspace of all possible flow rate variations $\delta \bm q_t$ that preserve mass conservation. The operator $\mathbbm{1}_t - H$ is precisely the orthogonal projector onto this null space. Thus, Eq.~\eqref{eq:jacobian_column_homogeneous} can be understood sequentially: the term $3a_0^{-1}q^0_{t,j}\bm{e}_j$ represents the linearised, ``naive'' local variation in the flow rate of the perturbed tube, respecting the Hagen-Poiseuille law in that specific tube while all other flow rates and pressures remain frozen. This naive variation is then projected by $\mathbbm{1}_t - H$ onto the subspace of physically admissible, mass-conserving flow changes. This projection effectively redistributes it across the network according to its topology.

The square of the Frobenius norm of $J^0$, which we require to compute the geometric contribution in Eq.~\eqref{eq:variances_relation}, is simply
the sum of the squares of the Euclidean norms of the column vectors defined by
Eq.~\eqref{eq:jacobian_column_homogeneous}. We obtain (see Appendix~\ref{app:computation_of_jacobian})
\begin{equation}
  \|J^0\|_\textrm{F}^2 = \frac{5}{3} N_p\frac{\overline{(q_{p}^0)^2}}{\langle A\rangle^2}.
  \label{eq:jacobian_frobenius_norm}
\end{equation}
Substituting this into~\eqref{eq:variances_relation} gives 
\begin{equation}
    \operatorname{CV}^2(Q_p) = \operatorname{CV}^2(Q_p^0) +\frac{5}{3}\left[1+\operatorname{CV}^2(Q_p^0)\right] \operatorname{CV}^2(A),
    \label{eq:variances_relation_final}
\end{equation}
where we have used
$ \overline{(q_{p}^0)^2}/ \overline{q_{p}^0}^2 = 1 +
\operatorname{CV}^2(Q_p^0)$.
Neglecting both instances of the topological contribution
$\operatorname{CV}^2(Q_p^0)$, which turns out to be small as already discussed, we obtain the surprisingly simple relationship
\begin{equation}
  \operatorname{CV}^2(Q_p) =  \frac{5}{3} \operatorname{CV}^2(A).
  \label{eq:variances_relation_simplified}
\end{equation}
This central result links the variability in flow rates to the variability in pore throat widths. Thus, although the functional shape of the flow distribution emerges from the network topology, its width is dominated by the geometry of the conduits. More specifically, the impact of throat size variability on flow rate statistics is found to be fully parameterisable in terms of a purely geometric measure that depends only on the first two statistical moments of the half-width distribution and does not require detailed knowledge of spatial distributions.

Having established this direct link between the flow rate variance and the geometry of the medium, we can now fully determine the flow statistics. Combining this result with the estimation of the mean pore flow rate from Section~\ref{sec:parameters:mean} allows us to evaluate $k$ and $\theta$ via equations~(\ref{eq:k_in_terms_of_statistics}) and~(\ref{eq:theta_in_terms_of_statistics}). Since these are the only two parameters on which the pore and tube PDFs, \eqref{eq:Pp} and \eqref{eq:Pt}, depend, both flow rate distributions are thus fully predictable from medium structure.

\section{Numerical validation}
\label{sec:validation}

In this section, we discuss the validation of the theoretical results in terms of both computational fluid dynamics (direct) numerical simulations and pore network simulations.

\subsection{Direct numerical simulations}

\begin{figure} 
    \centering
    \includegraphics[width=0.7\textwidth]{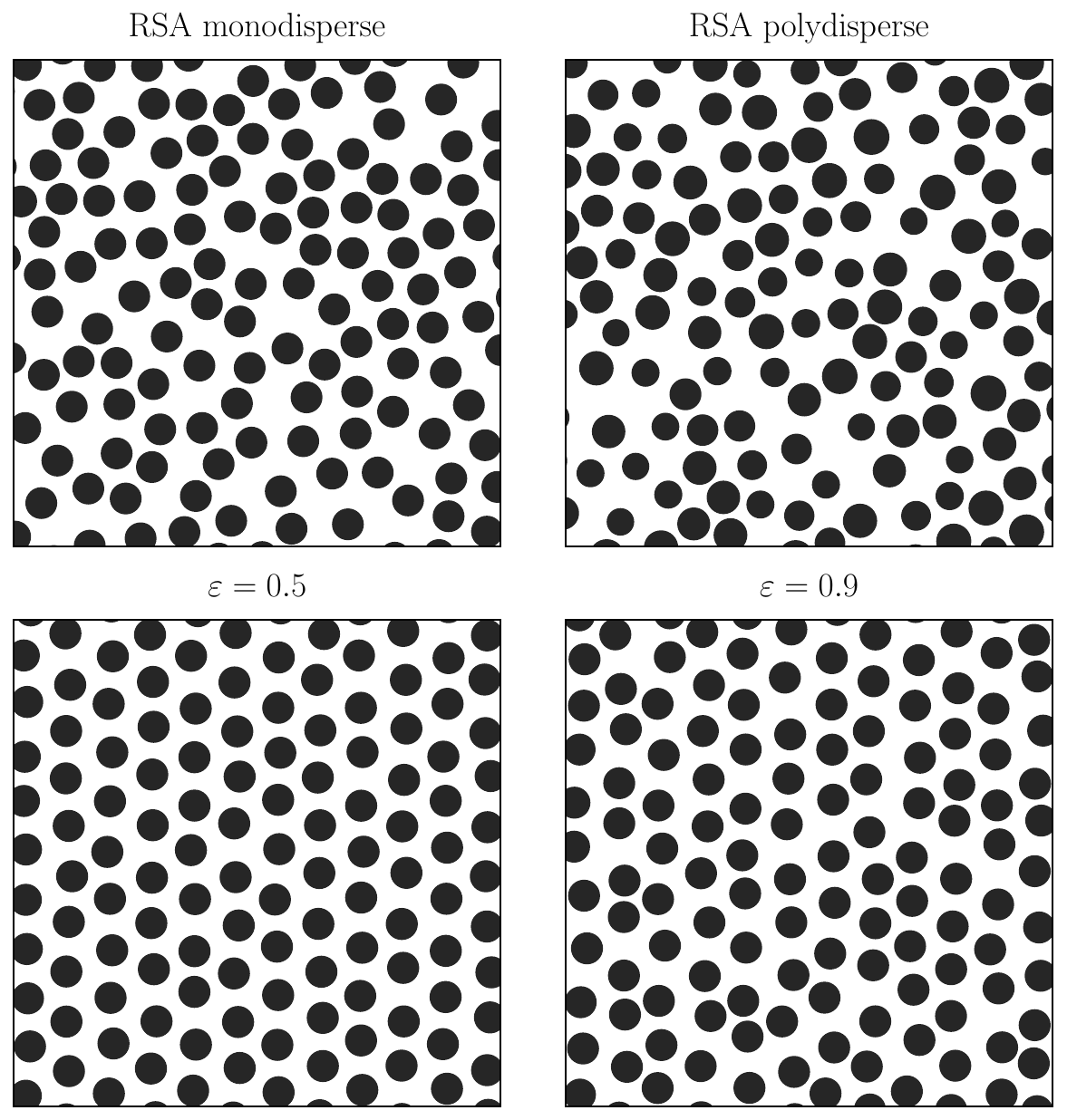}
    \caption{Zoomed-in views of some of the media. Note that the observation window spans $6.25\%$ of the total domain size to provide a clearer view of the disorder patterns typical of each medium.}
    \label{fig:medium_pictures}
\end{figure}

To test the validity of the \textit{Y} model, we perform direct numerical simulations in five distinct rectangular porous media (see figure~\ref{fig:medium_pictures}). Each medium has a length-to-width aspect ratio of $1.75$, and comprises $3120$ circular disks.
To induce spatial disorder in the positions of the disk centres, we employ two distinct generation algorithms: random sequential adsorption (RSA) \citep{talbot2000car}, and a randomly perturbed lattice approach \citep{khobaib2025gravity, pierce2026pore}, which we hereafter refer to as the $\varepsilon$ method. Both methods are detailed below. One of the five media is polydisperse, meaning its disk radii are non-uniform as explained subsequently. The other four media are monodisperse, characterised by a ratio of the disk radius to the domain width of $0.0083$. This value is adjusted to achieve a macroscopic porosity $\phi$ of approximately $0.6$ across all configurations. Flow in 2D systems is inherently more restricted than it is in 3D due to reduced spatial degrees of freedom. Furthermore, in RSA-generated media, the minimum achievable porosity is bounded to around 0.45 due to the so-called \textit{jamming limit} \citep{talbot2000car}. Thus, while this porosity is high when compared to those of natural 3D porous media, $\phi \simeq 0.6$ represents a relatively dense regime for a disordered 2D model, ensuring complex, tortuous flow while maintaining a connected pore network.

For the $\varepsilon$ method, the disks are initially placed on the sites of a regular triangular lattice with lattice spacing $b$. Spatial disorder is introduced by randomly displacing each disk centre. To guarantee that no disks overlap, the displacement vectors are generated by uniformly sampling random points within a circle of radius $\varepsilon(b/2 - r_0)$, where $r_0$ is the disk radius. The parameter $\varepsilon \in [0, 1]$ controls the amplitude of these random displacements, with $\varepsilon = 1$ representing the theoretical limit at which displaced adjacent disks could touch. This method allows for interpolating between ordered and disordered media, but the degree of disorder that can be attained is somewhat limited, as it is constrained by the initial regular grid.

To create disordered geometries that are not constrained by an underlying grid
while maintaining the same porosity, we employ an RSA algorithm \citep{talbot2000car}. In this method, disks are placed sequentially at uniformly random positions, rejecting overlaps. To avoid non-physical disk clustering and guarantee a realistic granular packing, we use an inflated exclusion radius for each circular obstacle. This radius is calculated so that the placement of exactly $3120$ identical disks corresponds to the system's jamming limit ($\phi\simeq0.453$). When this limit is reached, the radii of all placed disks are scaled down uniformly to yield the desired macroscopic porosity of $0.6$. 

Of the two RSA media so generated, one is kept monodisperse, while the other is made polydisperse by introducing a random variation to the individual disk radii. A naive approach to introduce this geometric disorder would be to sample the new radii from a uniform distribution bounded between $(1-\eta)r_0$ and $(1+\eta)r_0$, where $r_0$ is the original monodisperse radius and $\eta$ is a disorder control parameter. However, this procedure alters the expected area of the disks, thereby modifying the global porosity. To guarantee that the macroscopic porosity remains constant, we adopt a slightly different strategy: the maximum allowed radius is still set to $(1+\eta)r_0$, corresponding to an upper limit for the disk area of $\pi(1+\eta)^2 r_0^2$, but the new radii are generated such that the disk areas follow a uniform distribution that is centred at the original area $\pi r_0^2$. To maintain this mean area given this upper bound, the lower limit of the distribution must be $\pi [2 - (1+\eta)^2]r_0^2$, which corresponds to a minimum radius of $r_0\sqrt{2 - (1+\eta)^2}$. Note that this means that we must have $\eta < \sqrt{2} - 1 \simeq 0.414$ to avoid non-physical negative areas. In our polydisperse realisation, the disorder parameter was set to $\eta = 0.125$.

Using these media, we perform
direct numerical simulations of the two-dimensional Stokes equations for an incompressible fluid, 
\begin{subequations}
    \begin{equation}
        \mu\nabla^2\bm u = \bm \nabla p,
    \end{equation}
    \begin{equation}
        \bm \nabla \cdot \bm u = 0,
    \end{equation}
\end{subequations}
where $\bm u$ is the Eulerian velocity vector field, $p$ is the scalar pressure field, and $\mu$ is the dynamic viscosity of the fluid.

We apply no-slip boundary conditions on the disk surfaces and top and bottom medium edges,
and Dirichlet pressure conditions on the left and right medium boundaries. To compute the flow field, we employ a standard Navier-Stokes solver within the OpenFOAM framework \citep{weller1998tensorial, openfoam2406}. The dynamic viscosity $\mu$ is chosen such that the Reynolds number is small ($Re \simeq 4 \times 10^{-4}$), ensuring the flow remains well within the linear Stokes regime.

% We then apply statistical post-processing to the DNS results.
In each medium, we identify tubes and pores via Delaunay
triangulation, see Section~\ref{sec:theYmodel}. The flow rate through each tube
is calculated as the integral of the direct numerical simulation (DNS) velocity field over its cross-section,
and the flow rate of each pore is calculated as the sum of the flow rates of
the tubes for which it is the outlet pore. Up to small numerical precision
errors, the latter coincides with the sum of the
tube flow rates exiting the pore.

\begin{figure} 
    \centering \includegraphics[width=0.8\textwidth]{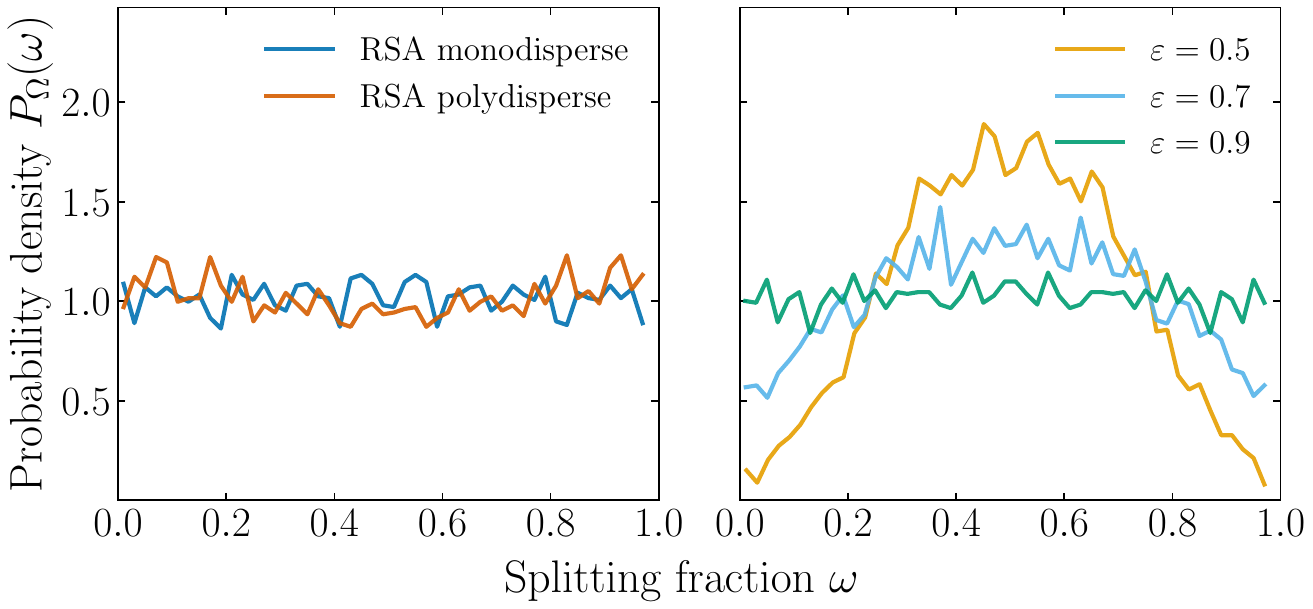}
    \caption{PDFs of the splitting fractions $\Omega$ at splitter junctions, extracted from direct numerical simulations. \textbf{(a)} As they exhibit a high level of spatial disorder, the RSA model fractions are well described by a uniform PDF. \textbf{(b)} In the $\varepsilon$ models, the PDF is highly sensitive to the magnitude of the random displacements. As the disorder parameter $\varepsilon$ increases, the initial peak at 0.5 continuously relaxes towards a uniform profile, effectively reaching it for the highest disorder case ($\varepsilon = 0.9$). 
    \label{fig:splitting_fractions}}
\end{figure}

Figure~\ref{fig:splitting_fractions} shows that for the RSA media and for the $\varepsilon=0.9$, the point statistics of splitting fractions $\Omega$ at splitter junctions (see figure~\ref{fig:delaunay_and_network}) are approximately uniform, in line with the theoretical developments in Section~\ref{sec:tube_flow_rate_distribution}.
Non-uniform splitting is observed only in exceptionally ordered geometries, at least for the range of media explored in this work (see figure~\ref{fig:medium_pictures}). Uniform splitting appears to be the typical behaviour, a feature that is also consistent with the observations reported by \citet{alim2017local}. In what follows, we focus on these media that exhibit approximately uniform splitting. The remaining media, which are highly ordered, are discussed in Appendix~\ref{app:ordered_limit}.

\begin{figure}
    \centering    \includegraphics[width=0.65\textwidth]{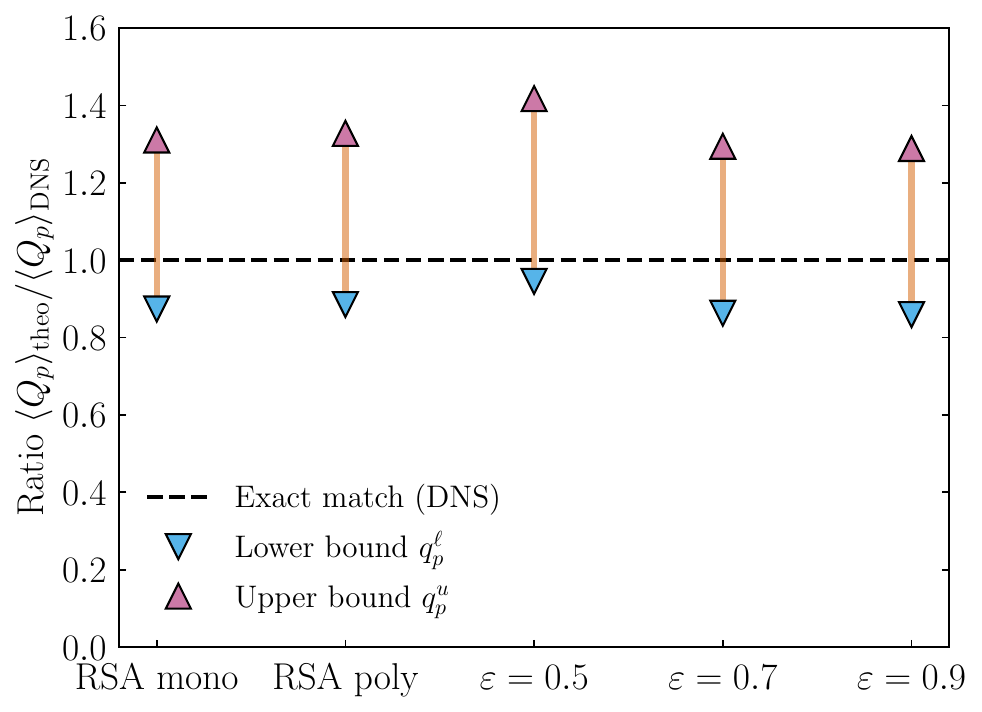}
    \caption{Comparison of the theoretical flow rate bounds with DNS data. The plot shows the ratio $\langle Q_p \rangle_{\mathrm{theo}} / \langle Q_p \rangle_{\mathrm{DNS}}$ for the lower ($\nabla$) and upper ($\Delta$) bounds, defined by Eqs.~\eqref{eq:lower_bound} and~\eqref{eq:upper_bound}.}
    \label{fig:mean_flow_rate}
\end{figure}

The mean flow rate predictions are shown in figure~\ref{fig:mean_flow_rate}. The true mean flow rate is bounded for all media by the limits~\eqref{eq:lower_bound} and~\eqref{eq:upper_bound} obtained from geometric arguments. Note that the lower bound~\eqref{eq:lower_bound} in particular always provides a reasonable estimate. The results that follow from here on are normalised by the corresponding mean flow rate, and therefore independent of it due to the linearity of Stokes flow. 

The \textit{Y} model predictions are in good agreement with the DNS flow rate distributions for both tubes and pores, with the Gamma shape parameter for the pore flow rates
%(equal to $1/\operatorname{CV}^2$)
derived directly from the geometric $\operatorname{CV}^2$ using the linear scaling of Eq.~\eqref{eq:variances_relation_simplified}. In particular, this approach significantly outperforms the mean-field approximation detailed in Appendix~\ref{app:mean_field_model}, see figure~\ref{fig:pore_and_throat_flow_rate_pdf}. We note that in the more ordered medium ($\varepsilon=0.9$), the predictions for the pore flow rate PDF perform more poorly. We ascribe this behaviour to the presence of longer-range correlations, which pose limits to the applicability of Lukacs's theorem. We note, however, that in the highly-ordered media corresponding to lower values of $\varepsilon$, the Gamma distribution again becomes a good description, see Appendix~\ref{app:ordered_limit}. At present, we lack a complete characterisation of when Lukacs's theorem can be expected to perform well in practice in quasi-ordered media, and whether these deviations disappear in the limit of infinite medium size.

\begin{figure}
    \centering    \includegraphics[width=\textwidth]{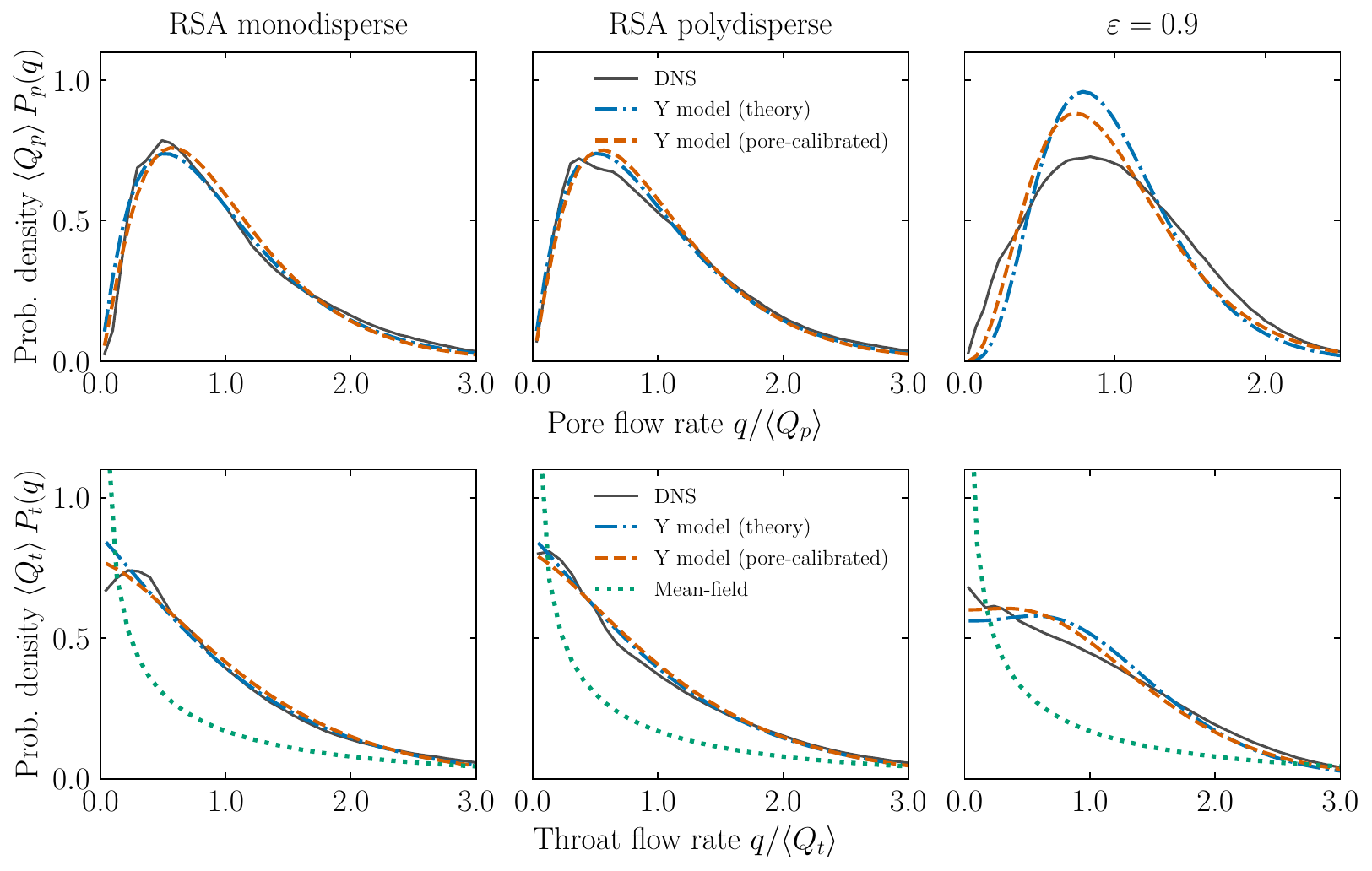}
    \caption{Comparison of the pore (top row) and throat (bottom row) flow rate distributions obtained from DNS against the analytical predictions of the \textit{Y} model and the mean-field model (detailed in Appendix~\ref{app:mean_field_model}) for the three most spatially disordered media.
    %Note that the mean-field prediction is only displayed for the throats, as this model does not account for flow distribution at the pore nodes.
    Because the flow variables are normalised by their respective means, the shape parameter $k$ of the pore flow Gamma distribution is the only governing parameter. For the dash-dotted blue lines (\textit{theory}), $k$ is predicted from the coefficient of variation of the geometric half-widths via Eqs.~\eqref{eq:variances_relation_simplified} and \eqref{eq:k_in_terms_of_statistics}. For the dashed red lines (\textit{pore-calibrated}), $k$ is fitted by matching the first two moments of the theoretical Gamma distribution to the DNS pore flow data. Once $k$ is fixed, the analytical pore and throat flow rate PDFs are determined by Eqs.~\eqref{eq:Pp} and \eqref{eq:Pt}, respectively.}
\label{fig:pore_and_throat_flow_rate_pdf}
\end{figure}

In any case, it should be noted that the standard notion of flow rate PDF, which considers all throats equally, corresponds to the throat flow rate PDF, as already discussed in section~\ref{sec:theYmodel}. For this metric, the \textit{Y} model performs reasonably well even in the latter case.
To better understand the strengths and limitations of the \textit{Y} model, including the small deviations observed for the throat flow rate distributions, we now turn to pore network simulations.

\subsection{Pore network simulations}

In addition to the direct numerical simulations of Stokes flow, we also carry
out pore network simulations. We construct the pore network
associated with each medium as discussed in Section~\ref{sec:theYmodel}. Then,
based on the direct simulation results, the effective geometric conductance of
each tube $j$ is calculated based on Eq.~\eqref{eq:Ohm} by dividing its flow rate by the modulus of the
pressure difference between its delimiting nodes $i_1$ and $i_2$: 
\begin{equation}
    c_{j} = \frac{\mu   q_{t,j}^{\textrm{DNS}}}{|p^{\textrm{DNS}}_{i_1} - p^{\textrm{DNS}}_{i_2}|} = C_{jj},
    \label{eq:conductance_list}
\end{equation}
where $c_{j}$ is the geometric conductance of the tube $j$ connecting pore $i_1$ to pore $i_2$, and corresponds to the $j$th entry of the diagonal of the conductance matrix $C$ in Eq.~\eqref{eq:Ohm},
$q_{t,j}^{\textrm{DNS}}$ is its computed flow rate, $\mu$ is the dynamic viscosity of the
fluid, and $p^\textrm{DNS}_{i_1}$ and $p^\textrm{DNS}_{i_2}$ are the pressure
values at the nodes. 

\begin{figure}
    \centering
    \includegraphics[width=\textwidth]{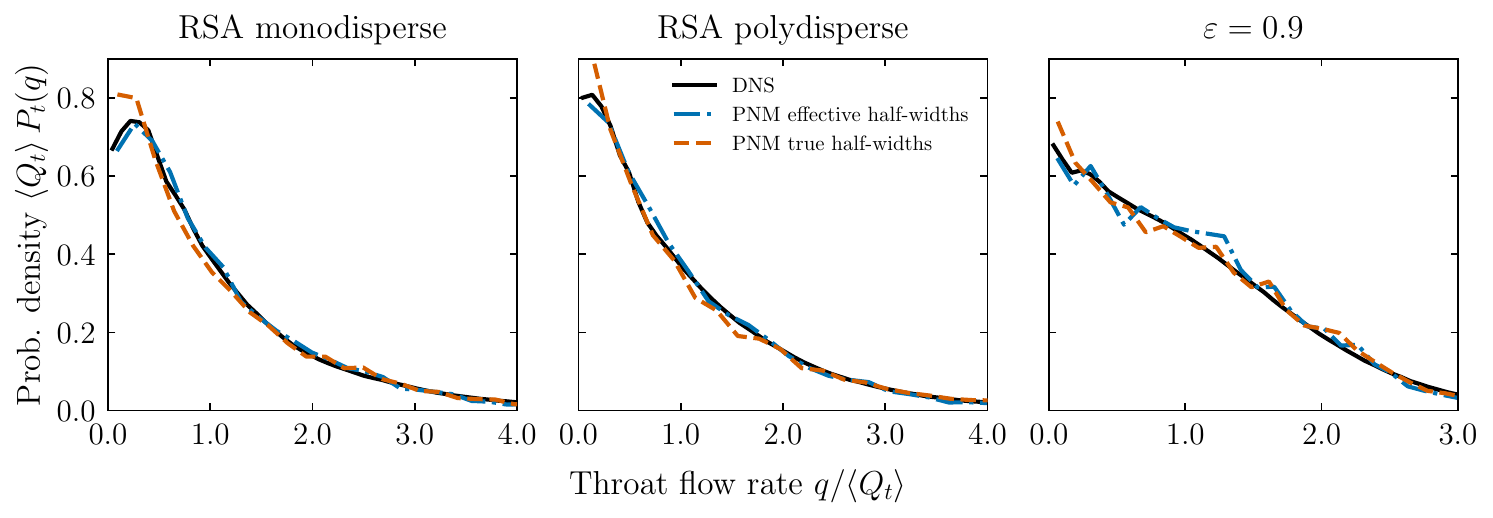}
    \caption{Validation of the pore network modelling (PNM) approach. The plot compares the scaled throat flow rate distributions obtained from: DNS (solid lines); a PNM simulation employing effective conductances derived from the DNS results (blue dash-dotted lines); and a PNM simulation using theoretical conductances based on a 2D Poiseuille tube model and the true measured half-widths of the medium (orange dashed lines).}
    \label{fig:pnm_validation} 
\end{figure}

Each PNM simulation consists of solving the linear system defined by Eqs.~\eqref{eq:KCL_Ohm} for the tube flow rates $\bm q_t$, using the geometric conductances from~\eqref{eq:conductance_list} and imposing the same pressure boundary conditions as in the DNS. Figure~\ref{fig:pnm_validation} shows that the flow rate distributions obtained from the PNM simulations agree closely with the direct numerical
simulations, validating the pore network approach.

\begin{figure}
    \centering
    \includegraphics[width=0.9\textwidth]{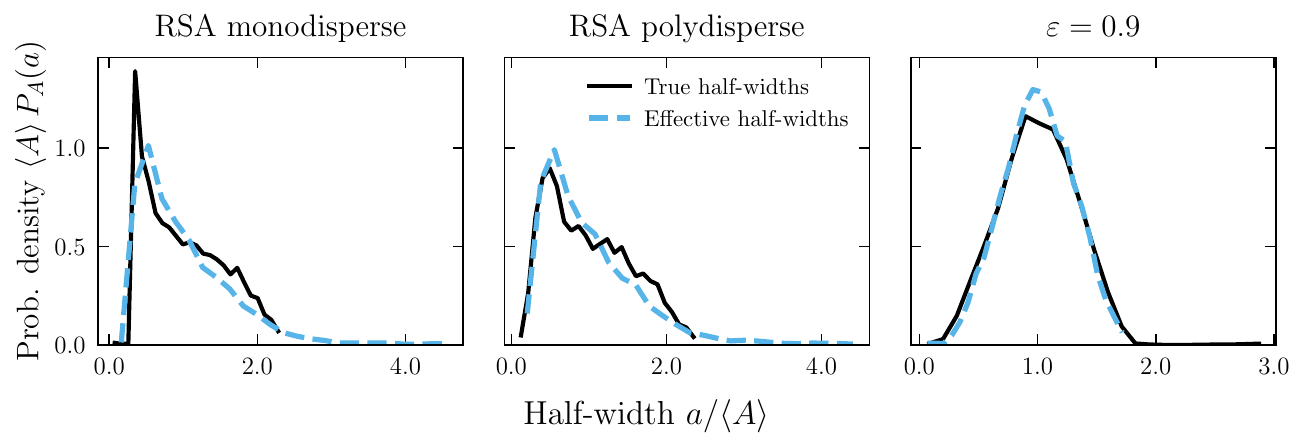}
    \caption{Comparison between the distributions of the actual geometric half-widths and the effective half-widths. The effective half-widths are obtained by inverting the Hagen-Poiseuille cubic law using the geometric conductances computed directly from the DNS, see Eq.~\eqref{eq:a_effective}. The PDFs of the normalised variables show good agreement, indicating that the underlying random variables differ only by a multiplicative constant.}
    \label{fig:half_width_comparison} 
\end{figure}

We now discuss the notion of \textit{true} and \textit{effective} half-widths.
Assuming a Poiseuille flow profile in each tube $j$, we can estimate its
effective half-width $a_{j}^{\textrm{eff}}$, i.e., the half-width of an idealised 2D tube
with the same conductance. It follows from the Hagen-Poiseuille law that the tube conductance is proportional to the cube of the half-width; inverting this relationship yields:
\begin{equation}
  \label{eq:a_effective}
    a_{j}^{\textrm{eff}} = \left( \frac{3}{2}   \ell_j   c_{j} \right)^{1/3},
\end{equation}
where $\ell_j$ is the length of the tube.
We find that the PDF of the normalised effective 
half-widths, $A^{\textrm{eff}} / \langle A^{\textrm{eff}} \rangle$, agrees closely 
with that of the true normalised half-widths, $A / \langle A \rangle$, see figure~\ref{fig:half_width_comparison}. 
This
% provides strong evidence
shows that the two variables differ 
essentially by a global multiplicative constant, $A^{\textrm{eff}} = \xi A$ for some $\xi>0$, which has important implications for the flow physics. 

In particular, if the geometric conductance of each tube is estimated from its true physical half-width using the standard cubic law, instead of from DNS flow and pressure values, the conductances will simply be uniformly scaled down 
by a factor of $\xi^3$. However, multiplying the conductance matrix $C$ by a scalar constant leaves the vector of 
flow rates $\bm{q}_t$ invariant, as shown by Eq.~\eqref{eq:tube_flows_in_terms_of_sources}. Consequently,
using the true geometric widths instead of the effective widths should recover the same flow statistics.

To verify this invariance principle, we compare two PNM simulations. First, we recall our previously validated PNM simulation, where we solve the flow field using the effective geometric conductances directly extracted from the DNS, as defined in Eq.~\eqref{eq:conductance_list}. Second, we re-compute the flow using geometric conductances computed from the true physical half-widths ($a_j$) via the cubic law. As expected, both approaches yield virtually identical flow rate PDFs (see figure~\ref{fig:pnm_validation}). This confirms that a simple conceptual tube model like the planar Poiseuille tube is sufficient to accurately capture the flow statistics, eliminating the need to compute effective conductances via DNS.

\begin{figure}
    \centering
    \includegraphics[width=\textwidth]{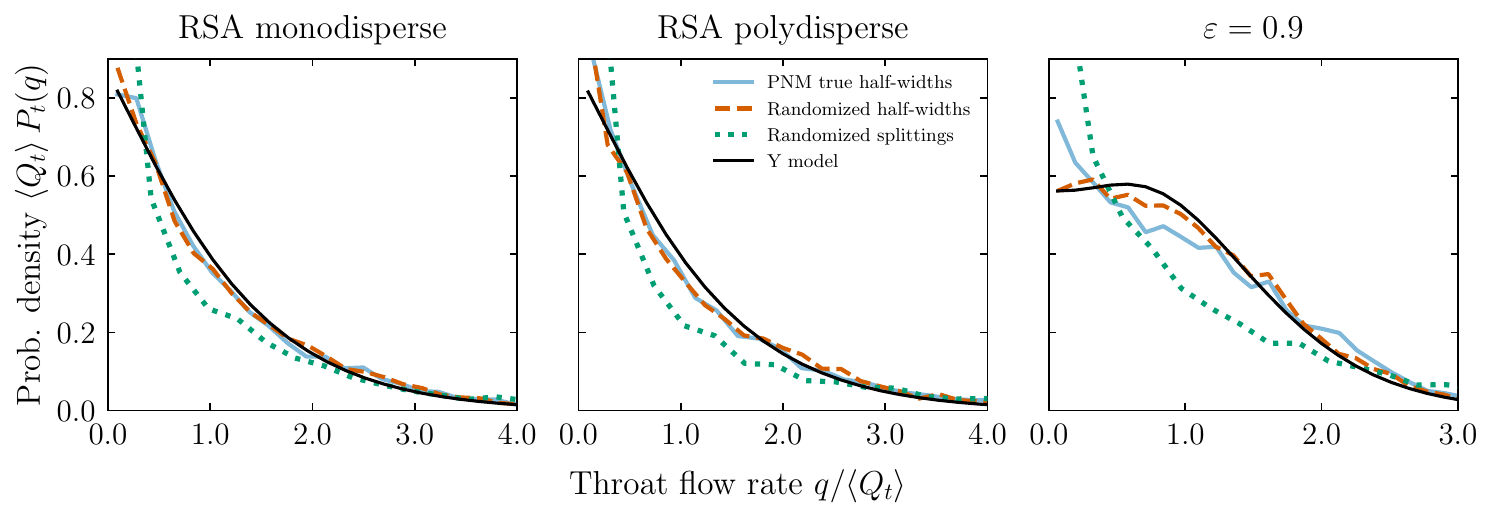}
    \caption{Comparison of the scaled throat flow rate distributions for the standard PNM (solid blue lines), a PNM with randomly shuffled half-widths (dashed orange lines), and a PNM with randomised splitting fractions (dotted green lines). The \textit{Y} model predictions (thin black lines) are included for reference; in the $\varepsilon=0.9$ case, it matches the PNM simulations with shuffled half-widths closer than the standard PNM simulations, suggesting that deviations from the theory in this more ordered case may be driven by spatial correlations.}
    \label{fig:randomized_models} 
\end{figure}

Having established that the true, physical half-widths are sufficient to capture the flow statistics, next we investigate the impact of spatial correlations between tube sizes. For this purpose, we perform additional PNM simulations where the true half-widths are randomly reshuffled across the existing network topology. We then calculate the new geometric conductance of each tube via the cubic law and solve for the flow field. Rather surprisingly, the resulting flow rate distribution is fundamentally unaffected by this removal of any spatial correlations, see figure~\ref{fig:randomized_models}. This demonstrates that, for the disordered media considered here, any existing local correlations between the sizes of neighbouring tubes are negligible regarding macroscopic flow statistics. The flow properties are instead primarily governed by the network topology and the point statistics of the geometry.

\begin{figure}
    \centering
    \includegraphics[width=0.61\textwidth]{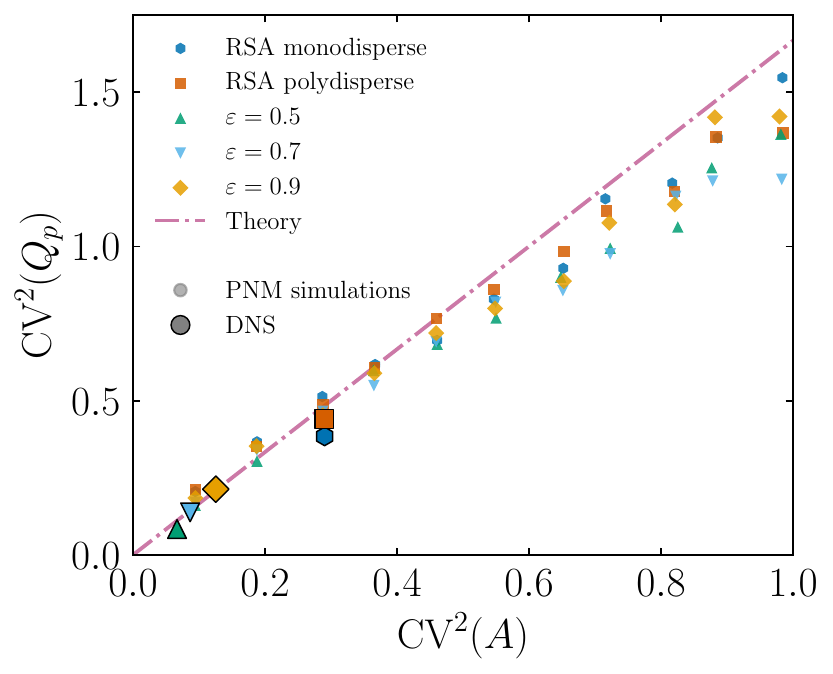}
    \caption{Comparison of the DNS and PNM simulation results against the theoretical scaling law~\eqref{eq:variances_relation_simplified}. The analytical prediction (dashed line) performs well even for media with higher tube size variability ($\mathrm{CV}^2(A) \simeq 1$). Large coloured symbols denote DNS results, while small coloured symbols denote PNM simulations with synthetic Gamma-sampled half-widths.} 
\label{fig:variance_scaling_law} 
\end{figure}

This insensitivity to spatial correlations allows us to easily generate synthetic geometries by independently sampling half-widths from any given distribution, providing a straightforward way to test our theoretical predictions. Specifically, we aim to validate the linear relationship between the coefficients of variation (CV) of the half-widths and the pore flow rates, as predicted by Eq.~\eqref{eq:variances_relation_simplified}. To do so, we  assign a ``synthetic", random half-width to each tube by sampling independently from a Gamma distribution. By fixing the mean half-width and modulating the shape parameter of the Gamma distribution, we smoothly vary the coefficient of variation of half-widths, which we have shown earlier to be a fundamental control on flow rate statistics. The PNM simulation solutions corroborate the linear behaviour~\eqref{eq:variances_relation_simplified} within its expected range of validity, see figure~\ref{fig:variance_scaling_law}.

One might wonder if spatial correlations between the splitting fractions defined in Section~\ref{sec:tube_flow_rate_distribution} are also negligible regarding flow statistics. As mentioned there, in sufficiently disordered media, the splitting fractions $\Omega$ at the flow splitters are approximately uniformly distributed, see figure~\ref{fig:splitting_fractions}. It is therefore tempting to independently sample these fractions $\Omega \sim U(0,1)$ at each flow splitter, without regard to the physical conductances, and then solve for the tube flow rates. However, randomising the splitting fractions fails to reproduce the correct flow statistics, yielding a flow rate PDF that deviates significantly from the standard PNM (see dotted lines in figure~\ref{fig:randomized_models}). Although any random set of splitting fractions yields a physical flow solution corresponding to some appropriate list of geometric conductances and obeying flow conservation, this associated geometry will generally have point statistics that differ drastically from those of the actual medium. Since the point statistics of the half-widths dictate the width of the flow distribution, preserving the true half-width distribution is critical. We thus conclude that, while the spatial correlations of the half-widths are largely irrelevant, spatial correlations in splitting fractions are fundamental for accurate flow statistics predictions because they
% encode the physics required to satisfy the correct 
arise as a direct consequence of geometric point statistics.

Two final observations regarding the results of the PNM simulations are in order. First, while the PNM employing DNS conductances accurately recreates the distribution at very low flow rates, the simulations relying on true physical half-widths (both standard and shuffled) exhibit a slight deviation from the DNS in this regime (see figures~\ref{fig:pnm_validation} and~\ref{fig:randomized_models}). This discrepancy suggests that the idealised Poiseuille tube model used in those simulations might not fully capture the hydrodynamics of tubes with very small flow rates, although we currently lack sufficient evidence to draw a definitive conclusion. Note that this deviation cannot be caused by spatial correlations, as the standard and shuffled Poiseuille networks coincide at low flows. Interestingly, the analytical \textit{Y} model predictions are closer to the Poiseuille-based PNM results than to the DNS in the low-flow regime. The precise reason for this remains unclear, in particular since this small discrepancy persists even when the shape parameter of the \textit{Y} model is directly fitted to the flow data rather than derived from the scaling law~\eqref{eq:variances_relation_simplified}, which is the only component of the model that explicitly assumes Poiseuille flow at the single-tube level.

Second, regarding the influence of spatial order, it is worth highlighting the $\varepsilon=0.9$ case (see figures~\ref{fig:pore_and_throat_flow_rate_pdf} and~\ref{fig:randomized_models}). In this medium, the theoretical prediction of the \textit{Y} model matches the PNM simulations with shuffled half-widths closer than it matches the standard PNM simulations. This suggests that the deviations from the theory observed in this more ordered medium are driven by spatial correlations that extend beyond the triplets of tubes at each \textit{Y}-shaped junction.

\section{Conclusions}
\label{sec:conclusion}

This work introduces the \textit{Y} model, a new pore-network-based approach to predict
flow rate statistics in disordered porous media that incorporates local flow correlations between
neighbouring pore throats.
The model predicts a Gamma distribution for the flow rates
through pore bodies via an application of Lukacs's theorem, and then uses
the statistical properties of \textit{Y}-shaped throat junctions to connect this
distribution to the flow rate statistics in pore throats. The resulting
predictions are in excellent agreement with both direct numerical and pore network simulations for disordered arrays of
two-dimensional disks generated according to different disorder models, for which
previous mean-field approaches break down.

The \textit{Y} model provides an analytical description of flow rate distributions through
pores and throats that is fully parameterised in terms of simple macroscopic
properties, namely the porosity, the size of the medium, and the average throat
width along with its coefficient of variation. In particular, we derive and
verify a surprisingly simple relationship connecting flow statistics to
geometry in disordered media: as long as the variability in throat sizes is
relatively mild, the coefficient of variation of flow through pore throats is
proportional to the coefficient of variation of throat widths. The
proportionality factor of $5/3$ appears to be universal for disordered 2D media, according to theory and simulations, as long as the underlying pore network is regular of degree 3 (see Section~\ref{sec:theYmodel}) and the Hagen-Poiseuille law applies in each throat. In particular, this relationship holds remarkably well for random arrays of disks, even when their
centres are qualitatively far from a regular lattice configuration. If either of the previous conditions breaks down, the proportionality constant could still be computed but would have a different value. 

While we conclude that topology-induced flow
correlations play an important role in flow statistics, we show numerically that independently resampling
the tube half-widths from their point distribution has little impact on the
global flow statistics. It is important to note that the original media and networks were generated using standard methods without any explicit constraint to suppress local correlations, so the observation that randomly resampling throat widths has no effect in flow statistics implies that local size correlations in these media are either naturally negligible or effectively irrelevant to the global flow. Thus, while previous work \citep{alim2017local} argues that local pore size correlations dictate macroscopic flow distributions in porous media, our findings indicate that in disordered media it is the underlying local topology (specifically, the \textit{Y}-shaped nature of the junctions) that fundamentally governs the shape of these distributions. In these media, only the width of the point statistics of throat apertures plays a significant role, by controlling the width of the flow distributions. Whether and to what extent media with strongly correlated structures break this invariance remains an open question for future work. 

Finally, while the \textit{Y} model focuses explicitly on predicting flow rate statistics, these results lay the background for determining the velocity PDFs required by stochastic transport models as discussed in the Introduction. Furthermore, although this study focuses only on fully-saturated conditions, the flow organisation into a backbone of preferential flow and low-velocity recirculation regions \citep{de1983hydrodynamic} in partially-saturated media can be exploited to extend saturated flow statistics models to these unsaturated conditions \citep{velasquez2022sharp}. While the low-flow regime statistics are driven by stagnation zones and are successfully captured by existing approaches, using the mean-field approximation for the backbone inaccurately represents the intermediate and high flow rates. Integrating the \textit{Y} model into this framework should improve the description of the complete flow rate spectrum.

The \textit{Y} model as presented here is restricted to two-dimensional porous media,
because it relies on the three-connected nature of planar graphs resulting from
disordered granular geometries. We believe, however, that its fundamental
building blocks, which rely on the statistical properties of
junctions rather than of single tubes, shed light on fundamental mechanisms
governing flow statistics under more general conditions. Whether and how these
principles generalise to 3D media and more complex geometries, including
fractured media, will be the subject of future work.

\section*{Funding}
The authors acknowledge financial support through the ERC Starting Grant project Uplift (101115760) funded by the EU. Views and opinions expressed are however those of the authors only and do not necessarily reflect those of the European Union or the European Research Council. Neither the European Union nor the granting authority can be held responsible for them. The authors also acknowledge financial support from MICIU/AEI/10.13039/501100011033 through Grant PID2024‐162869OB‐I00.

\section*{Declaration of Interests}
The authors report no conflict of interest.

\section*{Data availability statement}
The data that support the findings of this study are available from the corresponding author upon reasonable request.

\section*{Author ORCIDs}
J. Arnal, \url{https://orcid.org/0009-0002-4795-1615}; G. Sole-Mari, \url{https://orcid.org/0000-0002-9890-079X}; T. Aquino, \url{https://orcid.org/0000-0001-9033-7202}

\appendix

\section{Splitting and merging probabilities}
\label{app:probabilities}

In this appendix, we prove that in a regular pore network of degree three, the
probability $\beta$ that the inlet pore of a tube is a merger is $1/3$. First,
we consider the relationship between $\beta$ and the
probability $\alpha$ that a randomly chosen pore is a merger. Given any tube in
the network, its inlet pore is either a merger or a splitter. 
Each merger serves as the inlet for a single tube, while each splitter serves
as the inlet for two (see figure~\ref{fig:delaunay_and_network}). Therefore, if we count tubes by examining which pores
supply them with flow, each merger pore contributes one tube to the global
count, and each splitter pore contributes two. We can thus write the total
number $N_t$ of tubes in the system as:
\begin{equation}
  N_t = 2 N_s + N_m,
  \label{eq:tube_count_inletwise}
\end{equation}
where $N_m$ is the number of mergers and $N_s$ is the number of splitters. The
fraction of tubes having a merger as an inlet pore is thus
\begin{equation}
  \beta = \frac{N_m}{N_m + 2 N_s}
  % = \frac{N_m}{2N - N_m}
  = \frac{\alpha}{2 - \alpha},
  \label{eq:beta_in_terms_of_alpha}
\end{equation}
where we use $\alpha = N_m/N_p$, with $N_p = N_m + N_s$ the total number of pores.

Next, we find an alternative expression for the total number of tubes by
counting based on the outlet pore of each throat rather than the inlet. Again,
this outlet can be either a splitter or a merger.
% and these two situations are mutually exclusive.
Each merger serves as the outlet for two tubes, thus contributing two to the
total tube count, while each splitter serves as the outlet for one. Therefore,
we can write the total number of tubes in the network as:
\begin{equation}
  N_{t} = N_s + 2 N_m.
  \label{eq:tube_count_outletwise}
\end{equation}
Combining~\eqref{eq:tube_count_inletwise} and~\eqref{eq:tube_count_outletwise}, we find
% 
% \begin{equation}
$2 N_s + N_m = N_s + 2 N_m$.
% \end{equation}
We conclude that $N_s = N_m$, i.e., there are as many mergers as splitters.
Therefore, $\alpha = 1/2$, and substituting in
Eq.~\eqref{eq:beta_in_terms_of_alpha} we obtain
% 
% \begin{equation}
$\beta = 1/3$.
% \end{equation}

By assuming that the pore network representing the porous medium is regular of
degree three, we are implicitly neglecting the existence of boundary pores and thus assuming the medium is infinite or embedded in a toroidal
topology (i.e., periodic along both spatial directions). The pores at the inlet
and outlet of the medium have a connectivity lower than three, which breaks the
regularity of the network. The results $\alpha = 1/2$ and $\beta = 1/3$ are for
this reason applicable only if the number of boundary pores is
negligible compared to the number of bulk pores. Accounting for $N_i$ boundary inlet pores
and $N_o$ boundary outlet pores, Eqs.~\eqref{eq:tube_count_outletwise}
and~\eqref{eq:tube_count_inletwise} become
\begin{align}
  N_t = 2 N_s + N_m + N_i,
  &&N_t = N_s + 2 N_m + N_o.
  \label{eq:counting_with_borders}
\end{align}
Here we have assumed that all boundary pores have a connectivity of one. It is always possible to define the network such that this
is true: for instance, if a boundary inlet pore has degree 2, we can always add to the
network an auxiliary pore upstream that is connected to it through a single tube.
% The former inlet node becomes a bulk node, and the auxiliary node becomes a new
% inlet node with connectivity 1.
Equating the results in~\eqref{eq:counting_with_borders} one obtains
%
% \begin{equation}
$N_s + N_i = N_m + N_o$.
% \end{equation}
From these expressions and the fact that now $N_p = N_s + N_m + N_o + N_i$, it is
straightforward to derive the corrections to the values of probabilities
$\alpha$ and $\beta$ in the presence of boundary effects:
\begin{align}
  \alpha = \frac{1}{2} - P_o,
  &&\beta = \frac{1 - 2 P_o}{3 - P_i - P_o},
    \label{eq:corrected_alpha_beta}
\end{align}
where $P_o = N_o / N_p$ is the fraction of boundary outlet pores and $P_i = N_i / N_p$ is
the fraction of boundary inlet pores.

While $\alpha=1/2$ and $\beta=1/3$ are very good approximations for the media used in this work, an example of a regular network of degree 3 where the number of boundary
nodes is not negligible is obtained by taking the first $K$ shells of one of
the three main branches of a Bethe lattice (or Cayley tree) of degree 3 (for a clear visual representation see e.g. \citeauthor{brookings2005three}, \citeyear{brookings2005three}). For this network, $N_i=1$, $N_o=2^{K-1}$, and $N_p=2^K$. Therefore,
$P_o = 1/2$ and $P_i = 1/ 2^K $. Applying this to our corrected
formulas~\eqref{eq:corrected_alpha_beta} yields $\alpha = \beta = 0$, because in the Bethe lattice there are no merger pores.

\section{Lukacs's theorem and the Gamma distribution}
\label{app:lukacs_theorem}

The \textit{Y} model invokes Lukacs's proportion-sum independence theorem to justify the emergence of a Gamma distribution for the pore flow rates. Given its central importance, this appendix details this result and its extensions. From this theorem and other well-known properties of the Gamma distribution, a range of statistical properties of the flow rates can be deduced, which we summarise below.

\textbf{Theorem \citep[Lukacs's proportion-sum independence theorem,][]{lukacs1955characterization}.} Let $T_1$ and $T_2$ be two non-degenerate and positive random variables, and suppose that they are independently distributed. The random variables $Q_p = T_1 + T_2$ and $W_1 = T_1/Q_p$ are independently distributed if and only if both $T_1$ and $T_2$ have Gamma distributions with the same scale parameter.

\textbf{Corollary to Theorem 1 \citep{mosimann1962compound}}. Let $T_i$ ($i = 1, 2, \dots, N$) be $N$ non-degenerate and positive random variables, and suppose that they are independently distributed. Then each of the $N-1$ random variables $W_i = T_i/\sum_{j=1}^N  T_j$ ($i=1,\dots,N-1$) is distributed independently of the sum $Q_{p} = \sum_{j=1}^N T_j$ if and only if all $T_i$ have Gamma distributions with the same scale parameter.

\textbf{Theorem 2 \citep[distribution of fractions,][]{mosimann1962compound}.} Let $T_i$ ($i = 1, 2, \dots, N$) be $N$ independent, Gamma-distributed random variables with shape parameters $k_1, ..., k_N$ and equal scale parameter $\theta$. Then the vector of fractions $(W_1, \dots, W_N)$, with $W_i = T_i/\sum_{j=1}^N T_j$, follows a Dirichlet distribution with parameters $k_1, ..., k_N$.

\textbf{Property 1 (marginal distribution of fractions).} A direct consequence of the properties of a Dirichlet distribution with parameters $(k_1,...,k_N)$ is that the marginal distribution of any single fraction $W_i$ is a standard Beta distribution with parameters $k_i$ and $\sum_{j\neq i} k_j$. 

\textbf{Property 2 (sum of Gamma variables).} For a set of $N$ independent random variables $T_i$, each following a Gamma distribution with a shape parameter $k_i$ and a common scale parameter $\theta$, their sum $Q_{p} = \sum_{i=1}^N T_i$ is also Gamma-distributed with parameters $k =\sum_{i=1}^N k_i$ and $\theta$.

For the sake of completeness, we provide the explicit functional forms of the probability density functions discussed above. A positive, continuous random variable $Q_p$ follows a Gamma distribution with shape parameter $k > 0$ and scale parameter $\theta > 0$ if its probability density function (PDF) is given by

\begin{equation}
    P_p(q) = \frac{1}{\Gamma(k)\theta^k} q^{k-1} e^{-q/\theta}, \quad q > 0,
\end{equation}
where $\Gamma(\cdot)$ is the Gamma function. Similarly, a continuous random variable $W$ defined on the interval $(0, 1)$ follows a Beta distribution with shape parameters $b_1 > 0$ and $b_2 > 0$ if its PDF is
\begin{equation}
    P_W(w) = \frac{1}{B(b_1, b_2)} w^{b_1-1} (1-w)^{b_2-1}, \quad 0 < w < 1,
    \label{eq:beta_pdf}
\end{equation}
where $B(\cdot, \cdot)$ is the beta function. Finally, a random vector of proportions $\bm{W} = (W_1, \dots, W_N)$ constrained to the standard $(N-1)$-simplex (i.e., $W_i > 0$ and $\sum_{i=1}^N W_i = 1$) follows a Dirichlet distribution of order $N$ with concentration parameters $\bm{b} = (b_1, \dots, b_N)$ if its joint PDF is
\begin{equation}
P_{\bm{W}}(w_1, \dots, w_N) = \frac{\Gamma\left(\sum_{i=1}^N b_i\right)}{\prod_{i=1}^N \Gamma(b_i)} \prod_{i=1}^N w_i^{b_i-1}.
\end{equation}

In our theoretical framework, we use a combination of Theorem 1 (via its Corollary) and Property 2 to deduce that the pore flow rates are Gamma-distributed with parameters $k$ and $\theta$. However, the collection of results detailed above goes far beyond this single implication. From it, one can also deduce that each of the $N$ downstream partial flows $T_i=W_iQ_p$ is likewise Gamma-distributed with parameters $k/N$ and $\theta$. Consequently, the joint probability of the splitting fractions $(W_1, \dots, W_N)$ is characterised by a symmetric Dirichlet distribution with parameters ($k/N, ..., k/N$), and each individual fraction $W_i$ obeys a marginal Beta distribution with parameters $k/N$ and $(N-1)k/N$.

\section{Highly ordered limit}
\label{app:ordered_limit}

In highly ordered media, the uniform-splitting assumption detailed in Section~\ref{sec:tube_flow_rate_distribution} breaks down, see Figure~\ref{fig:splitting_fractions}. In this case, we find numerically that the PDF of the splitting fractions is well approximated by the more general symmetric Beta distribution, $\Omega \sim \B(b, b)$, of which the uniform distribution is a special case ($b=1$). Nevertheless, provided a minimal degree of structural disorder, the pore flow rates remain well approximated by a Gamma distribution as shown below.

To derive the tube flow rate PDF in this highly ordered configurations, we evaluate the probability density of the flow rates exiting a splitter pore, $Q_s = \Omega Q_p$. Using the symmetric Beta distribution, Eq.~\eqref{eq:beta_pdf} with $b_1 = b_2 = b$, and the Gamma distribution \eqref{eq:Pp} for the pore flow rates, the mixture integral \eqref{eq:pdf_of_product} becomes
\begin{equation}
    P_s(q) = \frac{\Gamma(2b)}{\Gamma(k) \theta^k \Gamma(b)^2} \int_q^\infty \frac{1}{q'} (q')^{k-1} e^{-q'/\theta} \left(\frac{q}{q'}\right)^{b-1} \left(1 - \frac{q}{q'}\right)^{b-1} dq'.
\end{equation}
Factoring out $q^{b-1}$ and introducing the change of variable $t = q' - q$ shifts the lower integration limit to zero:
\begin{equation}
    P_s(q) = \frac{\Gamma(2b) q^{b-1} e^{-q/\theta}}{\Gamma(k) \theta^k \Gamma(b)^2} \int_0^\infty t^{b-1} (t + q)^{k-2b} e^{-t/\theta}   dt.
\end{equation}
Next, we extract $q^{k-2b}$ from the binomial term and define a dimensionless variable $x = t/q$, which simplifies the integral to
\begin{equation}
    P_s(q) = \frac{\Gamma(2b) q^{k-1} e^{-q/\theta}}{\Gamma(k) \theta^k \Gamma(b)^2} \int_0^\infty x^{b-1} (1 + x)^{k-2b} e^{-q x/\theta}   dx.
\end{equation}
This matches the standard integral representation of Tricomi's confluent hypergeometric function, $U(a,c,z) = \Gamma(a)^{-1} \int_0^\infty x^{a-1} (1+x)^{c-a-1} e^{-zx}   dx$, with parameters $a = b$, $c = k - b + 1$, and $z = q/\theta$.
Substituting this result into the total tube flow rate distribution, $P_t(q) = P_p(q)/3 + 2P_s(q)/3$, leads to a somewhat more elaborate analytical expression than~\eqref{eq:Pt},
\begin{equation}
    P_t(q) = \frac{P_p(q)}{3} \left[ 1 + \frac{2\Gamma(2b)}{\Gamma(b)}   U\left(b,   k-b+1,   \frac{q}{\theta}\right) \right],
    \label{eq:Pt_generalized}
\end{equation}
where $P_p(q)$ is the Gamma distribution for pore flow rates defined in Eq.~\eqref{eq:Pp}.

\begin{figure}
    \centering
    \includegraphics[width=0.5\textwidth]{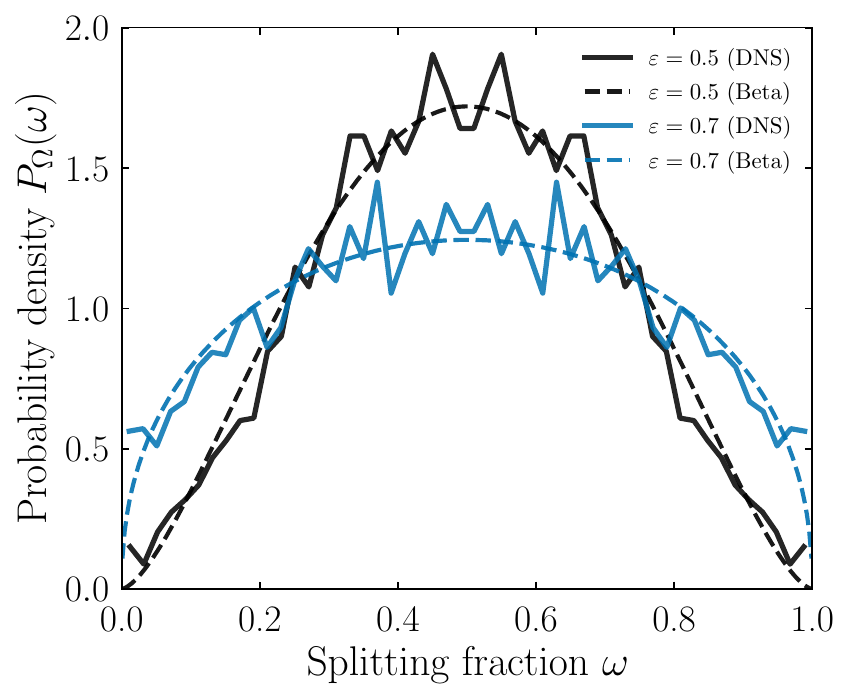}
    \caption{Empirical distributions of the splitting fraction $\Omega$ obtained from DNS (solid lines) for $\varepsilon = 0.5$ and $0.7$. Dashed lines correspond to symmetric Beta distributions, $\Omega \sim \B(b,b)$, with the shape parameter $b$ fitted via the method of moments.}
    \label{fig:splitting_fractions_ordered_limit} 
\end{figure}

Eq.~\eqref{eq:Pt_generalized} reduces to~\eqref{eq:Pt} for $b=1$ as expected. For arbitrary $b$, however, it is less predictive, as we currently lack an \textit{a priori} description of $b$ in terms of macroscopic medium properties.
Nevertheless, fitting $b$ by matching the second moment yields excellent agreement with the empirical splitting ratios (see figure~\ref{fig:splitting_fractions_ordered_limit}).

\begin{figure}
    \centering
    \includegraphics[width=0.99\textwidth]{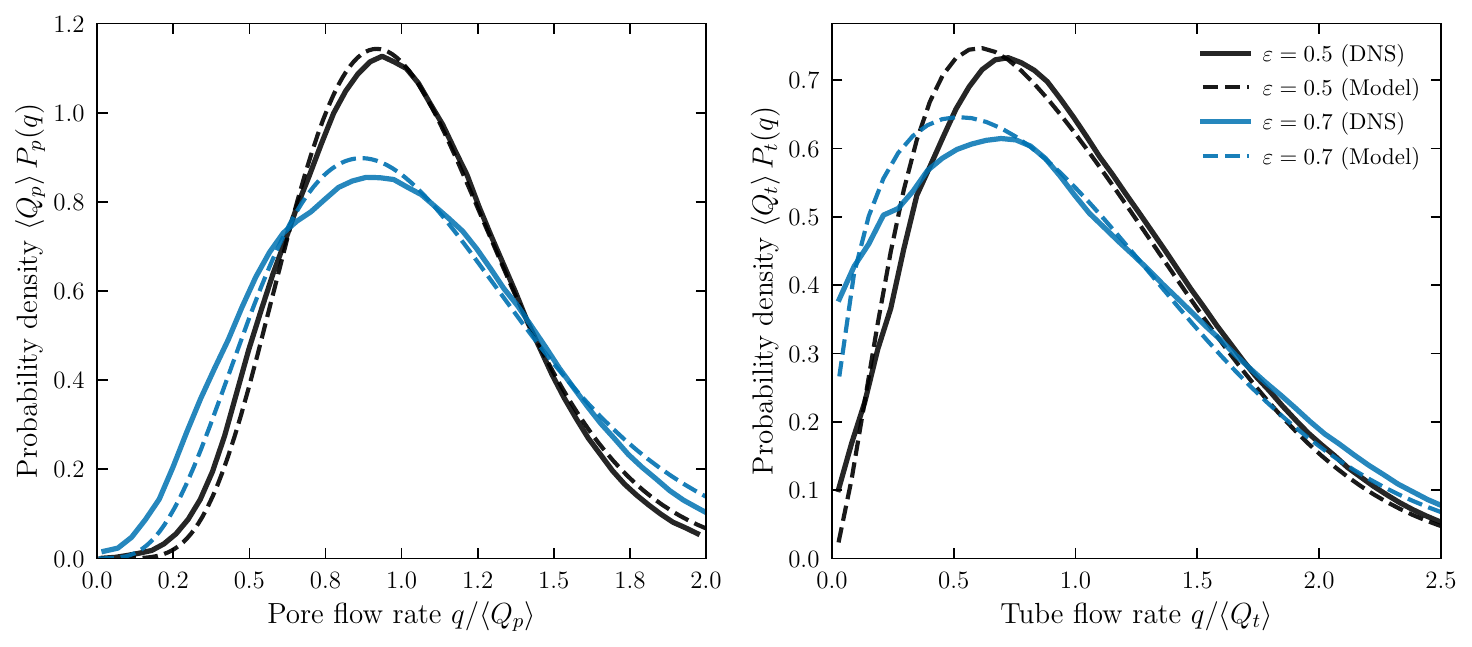}
    \caption{Probability density functions of normalised pore (left) and tube (right) flow rates for strongly ordered media ($\varepsilon = 0.5$ and $0.7$). Solid lines denote DNS results. Dashed lines represent the Gamma distribution fits for pores and the corresponding generalised analytical predictions for tubes, as given by Equation~(\ref{eq:Pt_generalized}). The analytical model shows good agreement with the empirical distributions.}
    \label{fig:flow_rates_ordered_limit} 
\end{figure}

Using the fitted splitting rates and fitted pore flow rate Gamma distributions as in the main text, we obtain a very good match for the pore throat flow rate distributions, see figure~\ref{fig:flow_rates_ordered_limit}. Note that, while here we obtain the pore Gamma distribution parameters through fitting because the goal is to illustrate the role of the splitting ratios, the relationship between the coefficient of variation of pore flow rates and of throat widths also holds for these media, see figure~\ref{fig:variance_scaling_law}. It could therefore be used for predictive determination of these parameters as before.

\section{Mean-field model}
\label{app:mean_field_model}

In this appendix, we briefly introduce the mean-field theory first presented by
\citet{coppersmith1996model} for distributions of forces in bead packs and then adapted by \citet{alim2017local} to flow distributions in porous media, and apply it to a disordered 2D collection of
circular obstacles to obtain the tube flow rate distribution. This
approximation assumes that the flow rates in the tubes are independent
realisations of the same random variable. Note that, although the \textit{Y} model also
treats tube flow rates as realisations of the same random variable, it does not
enforce their statistical independence. In fact, the \textit{Y} model predicts a strong
anticorrelation between the flow rates of two tubes emerging from the same
splitter, a phenomenon that we observe in direct simulations.

In the mean-field model, the only constraint on the flow rate probability
density function (PDF) is mass conservation at the inlet pore of each tube: 
\begin{equation}
    Q_{t,   i} = \sum_{j \in V_i} W_{ij} Q_{t, j}.
    \label{eq:mean_field_equation}
\end{equation}
That is, the flow rate of each tube $i$ is assumed to be a mixture of the flow
rates of the nearest upstream tubes $j$, which comprise the set $V_i$. These
flows are weighted by some splitting fractions $W_{ij}$, each of which is an
independent realisation of the same random variable $W$ with PDF $\eta(w)$. The
fractions $W_{ij}$ are also independent of the flow rates $Q_{t, j}$. It is
further assumed that all tubes have the same number $N$ of upstream neighbours.
Although this is an approximation, it is not as restrictive as it might appear,
since the splitting fractions $W_{ij}$ are allowed to be zero.

Note that while this formulation enforces Kirchhoff's first law at the inlet of each tube, it fails to guarantee mass conservation at the tube outlets. Specifically, the outgoing splitting fractions from a single upstream tube (which appear in separate instances of Eq.~\eqref{eq:mean_field_equation}) are not constrained to sum to unity. Instead, flow conservation at the tube outlets is only imposed statistically by enforcing that the average weight satisfies $\langle W \rangle = 1/N$.

Eq.~\eqref{eq:mean_field_equation} can be converted into a self-consistent
equation for $\tilde{P}_t(s)$, the Laplace transform of the tube flow rate
distribution,
\begin{equation}
\begin{split}
    \tilde{P}_t(s) &= \langle  e^{-s Q_t} \rangle 
    %= \langle e^{-s\sum_j w_j  Q_{t, j}} \rangle 
    = \prod_{j \in V_i}\langle e^{-s W_{ij} Q_{t, j}} \rangle = \langle e^{-s W Q_{t}} \rangle^N, \\
    &= \left[\int_0^1\int_0^\infty\eta(w)  e^{-swq} P_t(q) \textrm{d}q  \textrm{d}w \right]^N = \left[\int_0^1\eta(w) \tilde{P_t}(ws) \textrm{d}w \right]^N.
\end{split}
\label{eq:mean_field_laplace}
\end{equation}

As discussed in Section~\ref{sec:theYmodel}, in a disordered 2D collection of
disks all pores connect three tubes. Consequently, tubes have either one or two
upstream neighbours.
% ; no other configuration is possible.
However, we can consider that tubes with a single upstream neighbour (as seen in
figure~\ref{fig:delaunay_and_network}a) effectively have two, where one contributes a null fraction of its
flow ($w = 0$). We therefore set $N=2$.

We now proceed to derive the form of the fraction PDF $\eta(\omega)$. As
mentioned in Section~\ref{sec:theYmodel} and derived in
Appendix~\ref{app:probabilities}, $\beta=1/3$ of the tubes have a merger pore
as their inlet (figure~\ref{fig:delaunay_and_network}) while the remaining $2/3$ have a splitter. When a tube's inlet
is a merger, each of its two neighbours delivers its entire flow to it ($w=1$).
In contrast, when the inlet is a splitter, one of the two upstream
neighbours makes a null contribution ($w=0$), while the other contributes a
fraction of its flow which, as explained in Section~\ref{sec:theYmodel}, is
uniformly distributed between 0 and 1 across the ensemble of all splitter
nodes. Considering all these possible scenarios, the PDF of the splitting
fractions $w_j$ takes the form
\begin{equation}
  \eta(w) = \frac{1}{3}\delta(w-1) +\frac{2}{3}\left[\frac{1}{2} + \frac{1}{2} \delta(w)\right] = \frac{1}{3}\bigg[\delta(w-1) + 1 + \delta(w)\bigg],
  \label{eq:fraction_distribution}
\end{equation}
where $\delta(\cdot)$ is the Dirac delta.

Substituting this expression in
Eq.~\eqref{eq:mean_field_laplace}, we obtain
\begin{equation}
  \tilde{P_t}(s) = \frac{1}{9} \left[1 + \tilde{P_t}(s) + \int_0^1\tilde{P_t}(ws) \textrm{d}w\right]^2 .
\end{equation}
Defining $V(s) = \sqrt{\tilde{P}_t(s)}$ and taking the square root of both sides yields
\begin{equation}
  V(s) = \frac{1}{3} \left[1 + V(s) + \frac{1}{s}\int_0^sV^2(u) \textrm{d}u\right] ,
  \label{eq:equation_for_V}
\end{equation}
where the integral has been transformed using the change of variable $u = \omega s$. 

It is well known that the Laplace transform of a probability density function
is the moment-generating function with the sign of the argument reversed.
% % 
% \begin{equation}
%   \tilde{P_t}(s) = M_{Q_t}(-s).
% \end{equation}
Consequently, $\tilde{P}_t(0) = 1$, and the moments for $n\geqslant1$ obey
% % 
% \begin{equation}
%   \frac{\textrm{d}\tilde{P_t}(s)}{\textrm{d}s}\bigg|_{s=0} = -\langle Q_t \rangle   ,
% \end{equation}
% and, in general
% 
\begin{equation}
  \frac{\textrm{d}^n\tilde{P_t}(s)}{\textrm{d}s^n}\bigg|_{s=0} = (-1)^n \langle Q_t^n \rangle.
\end{equation}
Therefore, $V(0) = \sqrt{\tilde{P_t}(0)} = 1$ and
\begin{equation}
  \label{eq:mean_field_moments}
  \frac{\textrm{d}V(s)}{\textrm{d}s}\bigg|_{s=0} = \frac{\textrm{d}}{\textrm{d}s}\sqrt{\tilde{P_t}(s)}\bigg|_{s=0} = \frac{1}{2\sqrt{\tilde{P_t}(0)}} \frac{\textrm{d}\tilde{P_t}(s)}{\textrm{d}s}\bigg|_{s=0} = -\frac{1}{2} \langle Q_t \rangle.
\end{equation}
Multiplying both sides of Eq.~\eqref{eq:equation_for_V} by $s$ and subsequently differentiating with respect to $s$, we obtain
% , after rearranging terms:
% 
\begin{equation}
  \frac{1}{s} = \frac{3-2V}{1+2V^2-3V}   \frac{\textrm{d}V}{\textrm{d}s}.
\end{equation}
% The right-hand side can be decomposed into partial fractions,
% % 
% \begin{equation}
%   \frac{1}{s} = \left[\frac{1}{V-1} - \frac{4}{2V -1}\right]  \frac{\textrm{d}V}{\textrm{d}s}   .
% \end{equation}
Integrating both sides with respect to $s$ and exponentiation yields
\begin{equation}
  Cs = \frac{1-V}{(1-2V)^2}.
\end{equation}
The integration constant $C$ is obtained using~\eqref{eq:mean_field_moments} by
differentiating the previous expression with respect to $s$ and evaluating at
$s=0$, yielding $C = \langle Q_t \rangle/2$. Solving for $V$ from the previous equation and
squaring gives
% % 
% \begin{equation}
%   V(s) = \frac{1}{2} + \frac{\sqrt{1+4\langle Q_t\rangle s}-1}{4\langle  Q_t\rangle s}.
% \end{equation}
% Therefore,
% % 
\begin{equation}
  \tilde{P_t}(s) = \left(\frac{1}{2} + \frac{\sqrt{1+4\langle Q_t\rangle s}-1}{4\langle  Q_t\rangle s}\right)^2.
\end{equation}
The inverse Laplace transform of this expression can be computed using symbolic
calculation software. The exact mean-field expression for the tube flow rate
distribution in a disordered collection of disks becomes 
\begin{equation}
  P_t(q)
  = \frac{1}{4} \delta(q)
  + \frac{e^{-q/4\langle Q_t\rangle}}{2\sqrt{\pi \langle Q_t\rangle q}}
    \left(1 - \frac{q}{2\langle Q_t\rangle}\right)
  + \frac{q}{8 \langle Q_t\rangle^{2}} 
    \operatorname{erfc}\!\left(\frac{\sqrt{q}}{2\sqrt{\langle Q_t\rangle}}\right).
    \label{eq:mean_field_exact}
\end{equation}

A highly-accurate approximation of this exact analytical result can be obtained
via a simpler approach that also elucidates some properties of the mean-field
model. Note that although the partial flow rates $w_{ij}   Q_{t,j}$ are
independent by construction, Eq.~\eqref{eq:mean_field_equation} does not guarantee
that the total flow $Q_{t,i}$ and the splitting fractions $w_{ij}$ are
statistically independent. Given that the core premise of the mean-field approximation relies on the statistical decoupling of local variables, it is reasonable to ask whether there exists a distribution of fractions $\eta(w)$ that enforces this independence.
The answer is again provided by Lukacs's proportion-sum independence theorem:
given that the partial flow rates
% $\omega_j   Q_{t,j}$
are mutually independent, the total flow $Q_{t,i}$ and the fractions $w_{ij}$
are independent if and only if the flow rate distribution is Gamma. In that case, the
fractions follow the Beta distribution $\B[k/N, (N-1)k/N]$,
where $k$ is the shape parameter of the tube flow rate distribution, and $N$ is the
number of upstream neighbour tubes as before.

Although according to the arguments above the fraction distribution
$\eta(w)$ in the present case is not Beta but is rather given by
Eq.~\eqref{eq:fraction_distribution},
% almost
% any probability distribution defined on the interval $[0, 1]$ can be well
% approximated by a Beta distribution. Here, we attempt to approximate our
it can be approximated by the symmetric Beta distribution that shares its mean
and variance. Under this approximation, we write 
\begin{equation}
  \eta(w)  = \frac{\Gamma(2b)}{\Gamma^2(b)} w^{b-1}(1-w)^{b-1}
\end{equation}
for some $b>0$, with mean and variance $1/2$ and $1/(2b+1)$, respectively.
From~\eqref{eq:fraction_distribution}, we obtain $\langle W\rangle = 1/2$ and
$\operatorname{Var}(W)$ = 7/36, from which we conclude that $b = 1/7$.

\begin{figure}
    \centering
    \includegraphics[width=0.5\textwidth]{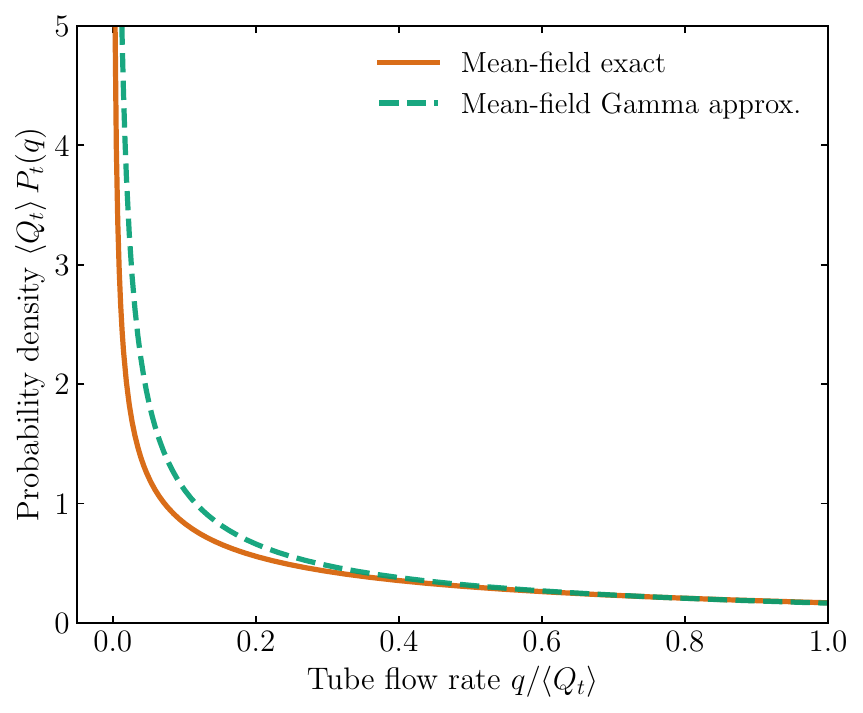}
    \caption{Analytical mean-field predictions for the normalised tube flow rate distribution in a disordered 2D array of disks. The solid line represents the exact theoretical solution derived by \citet{alim2017local}, while the dashed line corresponds to its corresponding Gamma approximation, presented in Eq.~\eqref{eq:mean_field_approx} .}
    \label{fig:alim_curves} 
\end{figure}

As previously mentioned, Lukacs's theorem connects the parameters of the Beta
fraction distribution with the shape parameter $k$ of the Gamma flow
distribution. In the present case, $k = 2b = 2/7$, so that
\begin{equation}
    P_t(q) = \frac{q^{-5/7} e^{-q/\theta}}{\Gamma(2/7) \theta^{2/7}},
    \label{eq:mean_field_approx}
\end{equation}
where the scale parameter $\theta$ is a characteristic flow rate given by
$\theta = \langle Q_t \rangle / k = 7\langle Q_t \rangle/2$. Figure~\ref{fig:alim_curves} shows that~\eqref{eq:mean_field_approx} provides a very good
approximation for~\eqref{eq:mean_field_exact}. This means that, although the exact solution of the mean-field model does not in fact yield tube flows $Q_{t,i}$ that are statistically independent of their upstream splitting fractions $w_{ij}$, it is remarkably close to the distribution that does. It should be recalled,
however, that both solutions differ substantially from those found from numerical
simulations, as discussed in Section~\ref{sec:validation}.

\section{Jacobian matrix of flow rates with respect to half-widths}
\label{app:computation_of_jacobian}

To derive the expression for the Jacobian $J^0$ presented in Eq.~\eqref{eq:jacobian_column_homogeneous}, we evaluate the change in pore flow rates induced by variations in tube half-widths around the homogeneous case ($\bm{a} = \bm{a}_0$). To do so, let us consider an infinitesimal change in the
conductance of the $j$th tube, $dc_j$, that induces a change $d\bm{q}_t$ in
the vector of tube flow rates, and a change $d\bm{q}_p$ in the vector of pore
flow rates. As both the source vector $\bm{s}$ and the incidence matrix $B$ are
constants with respect to the conductances, differentiation of
Eq.~\eqref{eq:KCL} with respect to $c_j$ leads to
\begin{equation}
    B \frac{\partial\bm{q}_t}{\partial c_j} = \bm{0},
\end{equation}
while differentiation of Eq.~\eqref{eq:Ohm} leads to
\begin{equation}
    \frac{\partial\bm{q}_t}{\partial c_j} = -\frac{\partial C}{\partial c_j}B^T\bm{p} - CB^T\frac{\partial\bm{p}}{\partial c_j}.
    \label{eq:ohm_differentiated}
\end{equation}
By definition, $c_j = C_{jj}$, that is, it is equal to the $j$th element of the diagonal of $C$, and thus $(\partial C/\partial c_j)_{ik} = \delta_{ik}\delta_{ij}$, where
$\delta_{\cdot\cdot}$ is the Kronecker delta (i.e., all entries are zero except
the one at position $(j,j)$, which is unity).

If we left-multiply Eq.~\eqref{eq:ohm_differentiated} by $B$, we obtain
\begin{equation*}
    \bm{0} = B\frac{\partial C}{\partial c_j}B^T\bm{p} + BCB^T\frac{\partial\bm{p}}{\partial c_j}.
\end{equation*}
We can thus solve for $\partial\bm{p}/\partial c_j$:
\begin{equation*}
    \frac{\partial\bm{p}}{\partial c_j} = -\left(BCB^T\right)^{-1}B\frac{\partial C}{\partial c_j}B^T\bm{p}.
\end{equation*}
Substituting this back into Eq.~\eqref{eq:ohm_differentiated} leads to
\begin{equation}
    \frac{\partial\bm{q}_t}{\partial c_j} = \left[\mathbbm{1}_t-CB^T\left(BCB^T\right)^{-1}B\right]\frac{\partial C}{\partial c_j}\Delta\bm{p},
\end{equation}
where $\mathbbm{1}_t$ is the identity matrix of rank $N_t$.
% However, it is not the derivative of $\bm{Q}_t$ with respect to $C_j$ we are interested in, but the one with respect to to the half-width $a_j$.
Using $c_j = 2a_j^3/(3 \ell_j)$, where $\ell_j$ is the length of tube $j$, this leads to
%it is easy to compute the latter from the former using the chain rule
%
\begin{equation}
    \frac{\partial \bm{q}_t}{\partial a_j} = \frac{\partial \bm{q}_t}{\partial c_j}\frac{2a_j^2}{\mu \ell_j} = \frac{\partial \bm{q}_t}{\partial c_j} \frac{3c_j}{a_j}.
\end{equation}
Therefore, the $j$th column of the Jacobian matrix reads
\begin{equation}
    \frac{\partial \bm{q}_p}{\partial a_j} =  \frac{3c_j}{a_j}Z\left[\mathbbm{1}_t-CB^T\left(BCB^T\right)^{-1}B\right]\frac{\partial C}{\partial c_j}\Delta\bm{p}.
    \label{eq:jacobian_column}
\end{equation}
% of the vector function $\bm{f}$ introduced in Eq.\eqref{eq:function_f}.
% We have thus effectively derived an expression for this Jacobian.

We now evaluate~\eqref{eq:jacobian_column} for the homogeneous case, that is, when
all tubes have the same half-width and therefore the same conductance (we
neglect the differences in length among tubes). In this case,
\begin{align}
  C = c_0\mathbbm{1}_t,
  &&\Delta\bm{p} = c_0^{-1} \bm{q}_t,
\end{align}
for some conductance $c_0$. Recalling that
$(\partial C/\partial c_j)_{ik} = \delta_{ik}\delta_{ij}$, we have
\begin{equation}
  \frac{\partial C}{\partial c_j}\Delta\bm{p} = \frac{1}{c_0}\frac{\partial C}{\partial c_j}\bm{q}_t = \frac{q_{t,j}}{c_0}\bm{e}_j,
\end{equation} 
where $\bm{e}_j$ is the $j$th vector of the standard basis of
$\mathbb{R}^{N_t}$. Substituting in~\eqref{eq:jacobian_column} leads directly to the final expression~\eqref{eq:jacobian_column_homogeneous} given in the main text.

The Frobenius norm of that result is
\begin{equation}
    \|J^0\|_\textrm{F}^2 =  \sum_{j=1}^{N_t}|3a_0^{-1}q^0_{t,j}Z(\mathbbm{1}_t-H)\bm{e}_j|^2 = a_0^{-2} \sum_{j=1}^{N_t}\lambda_j|q^0_{t,j}|^2,
    \label{eq:frobenius_norm}
\end{equation}
where $\lambda_j = |3Z(\mathbbm{1}_t-H)\bm{e}_j|^2$ is a measure of
how sensitive the pore flow rates are to a change in the half-width of tube
$j$. We now assume that there is no tube to which the network is especially
sensitive, which we expect to hold
%in particular
for disordered media where all
tubes have similar conductances and connectivity properties. Then, we
approximate the sensitivities by their average,
%
%\begin{equation}
$\lambda_j \simeq\bar{\lambda} = N_t^{-1}\sum_{k=1}^{N_t}\lambda_k$,
% \end{equation}
and Eq.~\eqref{eq:frobenius_norm} simplifies to
\begin{equation}
  \|J^0\|_{\textrm{F}}^2 = a_0^{-2}  N_t \bar{\lambda} \overline{(q_{t}^0)^2}.
    \label{eq:frobenius_norm_2}
\end{equation}
From the definition of the Euclidean norm, $\lambda_j = \sum_i|M_{ij}|^2$, where $M=3Z(\mathbbm{1}_t-H)$,
% $N_t \bar{\lambda}
% %=\sum_{j=1}^{N_t}|3Z(\mathbbm{1}_t-H)\bm{e}_j|^2
% = \sum_{i,j=1}^{N_t}|3[Z(\mathbbm{1}_t-H)]_{ij}|^2 = \|3Z(\mathbbm{1}_t-H)\|_\textrm{F}^2$,
so that, summing in $j$, we obtain
\begin{equation}
    N_t \bar{\lambda} 
    = \|3Z(\mathbbm{1}_t-H)\|_\textrm{F}^2
    = \operatorname{Tr}(9 Z\left[\mathbbm{1}_t-H\right]\left[\mathbbm{1}_t-H\right]^TZ^T) = 9 \operatorname{Tr}(Z\left[\mathbbm{1}_t-H\right]Z^T),
    \label{eq:trace}
\end{equation}
where we have used the fact that $\mathbbm{1}_t-H$ is an orthogonal projector.
By the cyclic property of the trace,
$\Tr( Z \left[\mathbbm{1}_t-H\right]  Z^T) = \Tr( Z^T Z
\left[\mathbbm{1}_t-H\right])$. Under the choice of labelling discussed in
Section~\ref{sec:parameters} for $Z$, $Z^T Z$ is an $N_t\times N_t$ diagonal
matrix with ones in its first $N_p$ diagonal entries and zeros elsewhere, and
thus the trace in Eq.~\eqref{eq:trace} is equal to the sum of the first $N_p$
diagonal elements of $\mathbbm{1}_t-H$. Assuming, similarly to above, that no
single diagonal entry of $\mathbbm{1}_t-H$ dominates the trace,
%, the value of the trace is roughly uniformly distributed along its diagonal, allowing us to
we approximate it as the full trace of $\mathbbm{1}_t-H$ scaled by $N_p/N_t$:
\begin{equation}
    N_t \bar{\lambda} \simeq 9 \frac{N_p}{N_t} \operatorname{Tr}(\mathbbm{1}_t-H) =9 \frac{N_p}{N_t}(N_t - N_p) = 3 N_p,
\end{equation}
where we have use the linearity of the trace and that
$\operatorname{Tr}(\mathbbm{1}_t) = N_t$, $\operatorname{Tr}(H) = N_p$ (because
it is a projector), and $N_p/N_t = 2/3$. Substituting in
Eq.~\eqref{eq:frobenius_norm_2} we obtain
\begin{equation}
  \|J^0\|_\textrm{F}^2 = 3 N_p\frac{\overline{(q_{t}^0)^2}}{\langle A\rangle^2}.
  \label{eq:jacobian_frobenius_norm_3}
\end{equation}

The quantity $\overline{(q_{t}^0)^2}$ appearing in the previous
expression is the second moment of the \textit{tube} flow rate distribution in
the homogeneous case. To relate it to the second moment of the \textit{pore}
flow rates, we can employ a topological double-counting argument. Let us
consider a typical pore where the total flow rate $q^0_{p,i}$ is carried by
a single incoming tube and splits into two outgoing branches (or vice versa).
As discussed previously, we can model this partition with a random splitting
fraction $\Omega\sim \mathcal{U}(0,1)$ that takes a value $\omega_i$ at node $i$, such that the outgoing (or incoming) flow rates are
$\omega_i q^0_{p,i}$ and $(1-\omega_i)q^0_{p,i}$. The sum of the
squared flow rates of the three tubes connected to pore $i$ is thus $(q^0_{p,i})^2 + (\omega_i   q^0_{p,i})^2 + [(1-\omega_i) q^0_{p,i}]^2$, and summing this over all $N_p$ pores in the network yields
\begin{equation}
% \begin{split}
    % \sum_{i=1}^{N_p}\left[(q^0_{p,i})^2 + (\omega_i   q^0_{p,i})^2 + [(1-\omega_i) q^0_{p}]^2\right]
    % &=
    N_p \overline{(q^0_{p})^2 + (\omega   q^0_{p})^2 + [(1-\omega) q^0_{p}]^2}
    % \\
    % &
    = N_p\left(1 + \frac{1}{3} + \frac{1}{3}\right)\overline{(q^0_{p})^2} =  \frac{5}{3} N_p\overline{(q^0_{p})^2}.
% \end{split}
\end{equation}
In this sum the flow rate of every tube is accounted for exactly twice (once at its inlet pore and once at its outlet pore), so that
%
% \begin{equation}
%  \frac{5}{3} N_p\overline{(q^0_{p})^2} 
%  =2\sum_{j=1}^{N_t}(q^0_{t,j})^2=2N_t\overline{(q^0_t)^2},
% \end{equation}
$5N_p\overline{(q^0_{p})^2}/3
 = 2N_t\overline{(q^0_t)^2}$, 
and thus, using $N_p/N_t=2/3$,
\begin{equation}
    \overline{(q_t^0)^2} = \frac{5}{9} \overline{(q_{p}^0)^2}.
\label{eq:second_moment_relation}
\end{equation}
Substituting~\eqref{eq:second_moment_relation} in~\eqref{eq:jacobian_frobenius_norm_3} yields the final result~\eqref{eq:jacobian_frobenius_norm} in the main text.

\bibliographystyle{unsrtnat}
\bibliography{references}

\begin{thebibliography}{60}
\expandafter\ifx\csname natexlab\endcsname\relax\def\natexlab#1{#1}\fi
\def\au#1{#1} \def\ed#1{#1} \def\yr#1{#1}\def\at#1{#1}\def\jt#1{\textit{#1}} \def\bt#1{#1}\def\bvol#1{\textbf{#1}} \def\vol#1{#1} \def\pg#1{#1} \def\publ#1{#1}\def\arxiv#1{#1}\def\org#1{#1}\def\st#1{\textit{#1}}

\bibitem[Alim {\em et~al.\/}(2017)Alim, Parsa, Weitz \& Brenner]{alim2017local}
{\sc \au{Alim, Karen}, \au{Parsa, Shima}, \au{Weitz, David~A} \& \au{Brenner, Michael~P}} \yr{2017}  \at{Local pore size correlations determine flow distributions in porous media}.  \jt{Physical Review Letters}  \bvol{119}~(14),  \pg{144501}.

\bibitem[de~Anna {\em et~al.\/}(2017)de~Anna, Quaife, Biros \& Juanes]{de2017prediction}
{\sc \au{de~Anna, Pietro}, \au{Quaife, Bryan}, \au{Biros, George} \& \au{Juanes, Ruben}} \yr{2017}  \at{Prediction of the low-velocity distribution from the pore structure in simple porous media}.  \jt{Physical Review Fluids}  \bvol{2}~(12),  \pg{124103}.

\bibitem[Aquino \& Le~Borgne(2021)]{aquino2021diffusing}
{\sc \au{Aquino, Tom{\'a}s} \& \au{Le~Borgne, Tanguy}} \yr{2021}  \at{The diffusing-velocity random walk: a spatial-markov formulation of heterogeneous advection and diffusion}.  \jt{Journal of Fluid Mechanics}  \bvol{910},  \pg{A12}.

\bibitem[Aurenhammer(1991)]{aurenhammer1991voronoi}
{\sc \au{Aurenhammer, Franz}} \yr{1991}  \at{Voronoi diagrams—a survey of a fundamental geometric data structure}.  \jt{ACM Computing Surveys (CSUR)}  \bvol{23}~(3),  \pg{345--405}.

\bibitem[Bear(2013)]{bear2013dynamics}
{\sc \au{Bear, Jacob}} \yr{2013} {\em Dynamics of fluids in porous media\/}.  \publ{Courier Corporation}.

\bibitem[Ben-Noah {\em et~al.\/}(2024)Ben-Noah, Hidalgo \& Dentz]{ben2024pore}
{\sc \au{Ben-Noah, Ilan}, \au{Hidalgo, Juan~J} \& \au{Dentz, Marco}} \yr{2024}  \at{Pore network models to determine the flow statistics and structural controls for single-phase flow in partially saturated porous media}.  \jt{Advances in Water Resources}  \bvol{193},  \pg{104809}.

\bibitem[Benson \& Cole(2008)]{benson2008co2}
{\sc \au{Benson, Sally~M} \& \au{Cole, David~R}} \yr{2008}  \at{{CO$_2$} sequestration in deep sedimentary formations}.  \jt{Elements}  \bvol{4}~(5),  \pg{325--331}.

\bibitem[Berkowitz {\em et~al.\/}(2006)Berkowitz, Cortis, Dentz \& Scher]{berkowitz2006modeling}
{\sc \au{Berkowitz, Brian}, \au{Cortis, Andrea}, \au{Dentz, Marco} \& \au{Scher, Harvey}} \yr{2006}  \at{Modeling non-fickian transport in geological formations as a continuous time random walk}.  \jt{Reviews of Geophysics}  \bvol{44}~(2).

\bibitem[Berkowitz \& Scher(1998)]{berkowitz1998theory}
{\sc \au{Berkowitz, Brian} \& \au{Scher, Harvey}} \yr{1998}  \at{Theory of anomalous chemical transport in random fracture networks}.  \jt{Physical Review E}  \bvol{57}~(5),  \pg{5858}.

\bibitem[Blunt(2017)]{blunt2017}
{\sc \au{Blunt, Martin~J}} \yr{2017} {\em Multiphase flow in permeable media: A pore-scale perspective\/}.  \publ{Cambridge University Press}.

\bibitem[Brookings {\em et~al.\/}(2005)Brookings, Carlson \& Doyle]{brookings2005three}
{\sc \au{Brookings, Ted}, \au{Carlson, J.~M.} \& \au{Doyle, John}} \yr{2005}  \at{Three mechanisms for power laws on the cayley tree}.  \jt{Physical Review E}  \bvol{72}~(5),  \pg{056120}.

\bibitem[Brutsaert(1966)]{brutsaert1966probability}
{\sc \au{Brutsaert, Wilfried}} \yr{1966}  \at{Probability laws for pore-size distributions}.  \jt{Soil Science}  \bvol{101}~(2),  \pg{85--92}.

\bibitem[Carman(1937)]{carman1937fluid}
{\sc \au{Carman, Philip~Crosbie}} \yr{1937}  \at{Fluid flow through granular beds}.  \jt{Trans. Inst. Chem. Eng. London}  \bvol{15},  \pg{150--156}.

\bibitem[Cieplak \& Robbins(1988)]{cieplak1988dynamical}
{\sc \au{Cieplak, Marek} \& \au{Robbins, Mark~O}} \yr{1988}  \at{Dynamical transition in quasistatic fluid invasion in porous media}.  \jt{Physical Review Letters}  \bvol{60}~(20),  \pg{2042}.

\bibitem[Coppersmith {\em et~al.\/}(1996)Coppersmith, Liu, Majumdar, Narayan \& Witten]{coppersmith1996model}
{\sc \au{Coppersmith, S.~N.}, \au{Liu, C.-H.}, \au{Majumdar, Satya}, \au{Narayan, Onuttom} \& \au{Witten, T.~A.}} \yr{1996}  \at{Model for force fluctuations in bead packs}.  \jt{Physical Review E}  \bvol{53}~(5),  \pg{4673}.

\bibitem[De~Gennes(1983)]{de1983hydrodynamic}
{\sc \au{De~Gennes, P.~G.}} \yr{1983}  \at{Hydrodynamic dispersion in unsaturated porous media}.  \jt{Journal of Fluid Mechanics}  \bvol{136},  \pg{189--200}.

\bibitem[Dentz \& de~Barros(2015)]{dentz2015mixing}
{\sc \au{Dentz, Marco} \& \au{de~Barros, Felipe P.~J.}} \yr{2015}  \at{Mixing-scale dependent dispersion for transport in heterogeneous flows}.  \jt{Journal of Fluid Mechanics}  \bvol{777},  \pg{178--195}.

\bibitem[Dentz {\em et~al.\/}(2004)Dentz, Cortis, Scher \& Berkowitz]{dentz2004time}
{\sc \au{Dentz, Marco}, \au{Cortis, Andrea}, \au{Scher, Harvey} \& \au{Berkowitz, Brian}} \yr{2004}  \at{Time behavior of solute transport in heterogeneous media: transition from anomalous to normal transport}.  \jt{Advances in Water Resources}  \bvol{27}~(2),  \pg{155--173}.

\bibitem[Dentz {\em et~al.\/}(2018)Dentz, Icardi \& Hidalgo]{dentz2018mechanisms}
{\sc \au{Dentz, Marco}, \au{Icardi, Matteo} \& \au{Hidalgo, Juan~J}} \yr{2018}  \at{Mechanisms of dispersion in a porous medium}.  \jt{Journal of Fluid Mechanics}  \bvol{841},  \pg{851--882}.

\bibitem[Dentz {\em et~al.\/}(2016)Dentz, Kang, Comolli, Le~Borgne \& Lester]{dentz2016continuous}
{\sc \au{Dentz, Marco}, \au{Kang, Peter~K}, \au{Comolli, Alessandro}, \au{Le~Borgne, Tanguy} \& \au{Lester, Daniel~R}} \yr{2016}  \at{Continuous time random walks for the evolution of lagrangian velocities}.  \jt{Physical Review Fluids}  \bvol{1}~(7),  \pg{074004}.

\bibitem[Dentz {\em et~al.\/}(2011)Dentz, Le~Borgne, Englert \& Bijeljic]{dentz2011mixing}
{\sc \au{Dentz, Marco}, \au{Le~Borgne, Tanguy}, \au{Englert, Andreas} \& \au{Bijeljic, Branko}} \yr{2011}  \at{Mixing, spreading and reaction in heterogeneous media: A brief review}.  \jt{Journal of Contaminant Hydrology}  \bvol{120},  \pg{1--17}.

\bibitem[Dullien(2012)]{dullien2012porous}
{\sc \au{Dullien, Francis A.~L.}} \yr{2012} {\em Porous media: fluid transport and pore structure\/}.  \publ{Academic press}.

\bibitem[Fatt(1956)]{fatt1956network}
{\sc \au{Fatt, Irving}} \yr{1956}  \at{The network model of porous media}.  \jt{Transactions of the AIME}  \bvol{207}~(01),  \pg{144--181}.

\bibitem[Freeze \& Cherry(1979)]{cherry1979groundwater}
{\sc \au{Freeze, R.~Allan} \& \au{Cherry, John~A.}} \yr{1979} {\em Groundwater\/}.  \publ{Englewood Cliffs, NJ: Prentice-Hall}.

\bibitem[Froment {\em et~al.\/}(1990)Froment, Bischoff \& De~Wilde]{froment1990chemical}
{\sc \au{Froment, Gilbert~F}, \au{Bischoff, Kenneth~B} \& \au{De~Wilde, Juray}} \yr{1990} {\em Chemical reactor analysis and design\/}, ,  \vol{vol.~2}.  \publ{Wiley New York}.

\bibitem[Golden(1980)]{golden1980percolation}
{\sc \au{Golden, J.~M.}} \yr{1980}  \at{Percolation theory and models of unsaturated porous media}.  \jt{Water Resources Research}  \bvol{16}~(1),  \pg{201--209}.

\bibitem[Hasimoto(1959)]{hasimoto1959periodic}
{\sc \au{Hasimoto, Hidenori}} \yr{1959}  \at{On the periodic fundamental solutions of the stokes equations and their application to viscous flow past a cubic array of spheres}.  \jt{Journal of Fluid Mechanics}  \bvol{5}~(2),  \pg{317--328}.

\bibitem[Heinemann {\em et~al.\/}(2021)Heinemann, Alcalde, Miocic, Hangx, Kallmeyer, Ostertag-Henning, Hassanpouryouzband, Thaysen, Strobel, Schmidt-Hattenberger {\em et~al.\/}]{heinemann2021enabling}
{\sc \au{Heinemann, Niklas}, \au{Alcalde, Juan}, \au{Miocic, Johannes~M}, \au{Hangx, Suzanne~JT}, \au{Kallmeyer, Jens}, \au{Ostertag-Henning, Christian}, \au{Hassanpouryouzband, Aliakbar}, \au{Thaysen, Eike~M}, \au{Strobel, Gion~J}, \au{Schmidt-Hattenberger, Cornelia} \& \au{others}} \yr{2021}  \at{Enabling large-scale hydrogen storage in porous media--the scientific challenges}.  \jt{Energy \& Environmental Science}  \bvol{14}~(2),  \pg{853--864}.

\bibitem[Herzig {\em et~al.\/}(1970)Herzig, Leclerc \& Le~Goff]{herzig1970flow}
{\sc \au{Herzig, John~Pierre}, \au{Leclerc, Denis~Maurice} \& \au{Le~Goff, P.}} \yr{1970}  \at{Flow of suspensions through porous media—application to deep filtration}.  \jt{Industrial \& Engineering Chemistry}  \bvol{62}~(5),  \pg{8--35}.

\bibitem[Ishino {\em et~al.\/}(2007)Ishino, Reyssat, Reyssat, Okumura \& Quere]{ishino2007wicking}
{\sc \au{Ishino, Chieko}, \au{Reyssat, Mathilde}, \au{Reyssat, Etienne}, \au{Okumura, Ko} \& \au{Quere, David}} \yr{2007}  \at{Wicking within forests of micropillars}.  \jt{EPL (Europhysics Letters)}  \bvol{79}~(5),  \pg{56005}.

\bibitem[Jerauld {\em et~al.\/}(1984)Jerauld, Hatfield, Scriven \& Davis]{jerauld1984percolation}
{\sc \au{Jerauld, G.~R.}, \au{Hatfield, J.~C.}, \au{Scriven, L.~E.} \& \au{Davis, H.~T.}} \yr{1984}  \at{Percolation and conduction on voronoi and triangular networks: a case study in topological disorder}.  \jt{Journal of Physics C: Solid State Physics}  \bvol{17}~(9),  \pg{1519--1529}.

\bibitem[Kang {\em et~al.\/}(2011)Kang, Dentz, Le~Borgne \& Juanes]{kang2011spatial}
{\sc \au{Kang, Peter~K}, \au{Dentz, Marco}, \au{Le~Borgne, Tanguy} \& \au{Juanes, Ruben}} \yr{2011}  \at{Spatial markov model of anomalous transport through random lattice networks}.  \jt{Physical Review Letters}  \bvol{107}~(18),  \pg{180602}.

\bibitem[Kerstein(1983)]{kerstein1983equivalence}
{\sc \au{Kerstein, Alan~R}} \yr{1983}  \at{Equivalence of the void percolation problem for overlapping spheres and a network problem}.  \jt{Journal of Physics A: Mathematical and General}  \bvol{16}~(13),  \pg{3071--3075}.

\bibitem[Khobaib {\em et~al.\/}(2025)Khobaib, Reis, Moura, Toussaint, Flekk{\o}y \& M{\aa}l{\o}y]{khobaib2025gravity}
{\sc \au{Khobaib, Khobaib}, \au{Reis, Paula}, \au{Moura, Marcel}, \au{Toussaint, Renaud}, \au{Flekk{\o}y, Eirik~Grude} \& \au{M{\aa}l{\o}y, Knut~J{\o}rgen}} \yr{2025}  \at{Gravity stabilized drainage in porous media with controlled disorder}.  \jt{Physical Review Research}  \bvol{7}~(2),  \pg{023040}.

\bibitem[Kosugi(1994)]{kosugi1994three}
{\sc \au{Kosugi, Ken'ichirou}} \yr{1994}  \at{Three-parameter lognormal distribution model for soil water retention}.  \jt{Water Resources Research}  \bvol{30}~(4),  \pg{891--901}.

\bibitem[Kozeny(1927)]{kozeny1927ueber}
{\sc \au{Kozeny, Josef}} \yr{1927}  \at{Ueber kapillare leitung des wassers im boden}.  \jt{Sitzungsberichte der Akademie der Wissenschaften in Wien}  \bvol{136},  \pg{271}.

\bibitem[Kutsovsky {\em et~al.\/}(1996)Kutsovsky, Scriven, Davis \& Hammer]{kutsovsky1996nmr}
{\sc \au{Kutsovsky, Y.~E.}, \au{Scriven, L.~E.}, \au{Davis, H.~T.} \& \au{Hammer, B.~E.}} \yr{1996}  \at{Nmr imaging of velocity profiles and velocity distributions in bead packs}.  \jt{Physics of Fluids}  \bvol{8}~(4),  \pg{863--871}.

\bibitem[Le~Borgne {\em et~al.\/}(2008)Le~Borgne, Dentz \& Carrera]{le2008lagrangian}
{\sc \au{Le~Borgne, Tanguy}, \au{Dentz, Marco} \& \au{Carrera, Jesus}} \yr{2008}  \at{Lagrangian statistical model for transport in highly heterogeneous velocity fields}.  \jt{Physical Review Letters}  \bvol{101}~(9),  \pg{090601}.

\bibitem[Liu {\em et~al.\/}(2026)Liu, Wang \& Wang]{liu2026mechanism}
{\sc \au{Liu, Yang}, \au{Wang, Yuedi} \& \au{Wang, Moran}} \yr{2026}  \at{Mechanism transition of superlinear scaling in hydrodynamic dispersion}.  \jt{Journal of Fluid Mechanics}  \bvol{1031},  \pg{A32}.

\bibitem[Liu {\em et~al.\/}(2024)Liu, Xiao, Aquino, Dentz \& Wang]{liu2024scaling}
{\sc \au{Liu, Yang}, \au{Xiao, Han}, \au{Aquino, Tom{\'a}s}, \au{Dentz, Marco} \& \au{Wang, Moran}} \yr{2024}  \at{Scaling laws and mechanisms of hydrodynamic dispersion in porous media}.  \jt{Journal of Fluid Mechanics}  \bvol{1001},  \pg{R2}.

\bibitem[Lukacs(1955)]{lukacs1955characterization}
{\sc \au{Lukacs, Eugene}} \yr{1955}  \at{A characterization of the gamma distribution}.  \jt{The Annals of Mathematical Statistics}  \bvol{26}~(2),  \pg{319--324}.

\bibitem[Manthiram {\em et~al.\/}(2014)Manthiram, Fu, Chung, Zu \& Su]{manthiram2014}
{\sc \au{Manthiram, Arumugam}, \au{Fu, Yongzhu}, \au{Chung, Sheng-Heng}, \au{Zu, Chenxi} \& \au{Su, Yu-Sheng}} \yr{2014}  \at{Rechargeable lithium--sulfur batteries}.  \jt{Chemical Reviews}  \bvol{114}~(23),  \pg{11751--11787}.

\bibitem[Mosimann(1962)]{mosimann1962compound}
{\sc \au{Mosimann, James~E.}} \yr{1962}  \at{On the compound multinomial distribution, the multivariate $\beta$-distribution, and correlations among proportions}.  \jt{Biometrika}  \bvol{49}~(1/2),  \pg{65--82}.

\bibitem[{OpenCFD Ltd}(2024)]{openfoam2406}
{\sc \au{{OpenCFD Ltd}}} \yr{2024} {OpenFOAM} -- the open source {CFD} toolbox, version 2406. \url{https://www.openfoam.com}.

\bibitem[Orr~Jr. \& Taber(1984)]{orr1984use}
{\sc \au{Orr~Jr., F.~M.} \& \au{Taber, J.~J.}} \yr{1984}  \at{Use of carbon dioxide in enhanced oil recovery}.  \jt{Science}  \bvol{224}~(4649),  \pg{563--569}.

\bibitem[Pierce {\em et~al.\/}(2026)Pierce, Le~Borgne, Renard \& Linga]{pierce2026pore}
{\sc \au{Pierce, J.~Kevin}, \au{Le~Borgne, Tanguy}, \au{Renard, Francois} \& \au{Linga, Gaute}} \yr{2026}  \at{How pore-scale disorder controls fluid stretching in porous media}.  \jt{arXiv preprint arXiv:2604.02981} .

\bibitem[Purcell(1949)]{purcell1949capillary}
{\sc \au{Purcell, W.~R.}} \yr{1949}  \at{Capillary pressures-their measurement using mercury and the calculation of permeability therefrom}.  \jt{Journal of Petroleum Technology}  \bvol{1}~(02),  \pg{39--48}.

\bibitem[Reis {\em et~al.\/}(2023)Reis, Moura, Linga, Rikvold, Toussaint, Flekk{\o}y \& M{\aa}l{\o}y]{reis2023simplified}
{\sc \au{Reis, Paula}, \au{Moura, Marcel}, \au{Linga, Gaute}, \au{Rikvold, Per~Arne}, \au{Toussaint, Renaud}, \au{Flekk{\o}y, Eirik~Grude} \& \au{M{\aa}l{\o}y, Knut~J{\o}rgen}} \yr{2023}  \at{A simplified pore-scale model for slow drainage including film-flow effects}.  \jt{Advances in Water Resources}  \bvol{182},  \pg{104580}.

\bibitem[Sahimi(2011)]{sahimi2011flow}
{\sc \au{Sahimi, Muhammad}} \yr{2011} {\em Flow and transport in porous media and fractured rock: from classical methods to modern approaches\/}.  \publ{John Wiley \& Sons}.

\bibitem[Sangani \& Acrivos(1982)]{sangani1982slow}
{\sc \au{Sangani, Ashok~Shantilal} \& \au{Acrivos, Andreas}} \yr{1982}  \at{Slow flow past periodic arrays of cylinders with application to heat transfer}.  \jt{International Journal of Multiphase Flow}  \bvol{8}~(3),  \pg{193--206}.

\bibitem[Sangani \& Yao(1988)]{sangani1988transport}
{\sc \au{Sangani, Ashok~S} \& \au{Yao, C}} \yr{1988}  \at{Transport processes in random arrays of cylinders. ii. viscous flow}.  \jt{The Physics of Fluids}  \bvol{31}~(9),  \pg{2435--2444}.

\bibitem[Shannon {\em et~al.\/}(2008)Shannon, Bohn, Elimelech, Georgiadis, Mari{\~n}as \& Mayes]{shannon2008}
{\sc \au{Shannon, Mark~A}, \au{Bohn, Paul~W}, \au{Elimelech, Menachem}, \au{Georgiadis, John~G}, \au{Mari{\~n}as, Benito~J} \& \au{Mayes, Anne~M}} \yr{2008}  \at{Science and technology for water purification in the coming decades}.  \jt{Nature}  \bvol{452}~(7185),  \pg{301--310}.

\bibitem[Sherman {\em et~al.\/}(2021)Sherman, Engdahl, Porta \& Bolster]{sherman2021review}
{\sc \au{Sherman, Thomas}, \au{Engdahl, Nicholas~B}, \au{Porta, Giovanni} \& \au{Bolster, Diogo}} \yr{2021}  \at{A review of spatial markov models for predicting pre-asymptotic and anomalous transport in porous and fractured media}.  \jt{Journal of Contaminant Hydrology}  \bvol{236},  \pg{103734}.

\bibitem[Szulczewski {\em et~al.\/}(2012)Szulczewski, MacMinn, Herzog \& Juanes]{szulczewski2012lifetime}
{\sc \au{Szulczewski, Michael~L}, \au{MacMinn, Christopher~W}, \au{Herzog, Howard~J} \& \au{Juanes, Ruben}} \yr{2012}  \at{Lifetime of carbon capture and storage as a climate-change mitigation technology}.  \jt{Proceedings of the National Academy of Sciences}  \bvol{109}~(14),  \pg{5185--5189}.

\bibitem[Talbot {\em et~al.\/}(2000)Talbot, Tarjus, Van~Tassel \& Viot]{talbot2000car}
{\sc \au{Talbot, J.}, \au{Tarjus, G.}, \au{Van~Tassel, P.~R.} \& \au{Viot, P.}} \yr{2000}  \at{From car parking to protein adsorption: an overview of sequential adsorption processes}.  \jt{Colloids and Surfaces A: Physicochemical and Engineering Aspects}  \bvol{165}~(1-3),  \pg{287--324}.

\bibitem[{United Nations}(2022)]{UnitedNations2022}
{\sc \au{{United Nations}}} \yr{2022} {\em The United Nations World Water Development Report 2022: Groundwater: Making the invisible visible\/}.  \publ{Paris: UNESCO}.

\bibitem[Valvatne \& Blunt(2004)]{valvatne2004predictive}
{\sc \au{Valvatne, Per~H} \& \au{Blunt, Martin~J}} \yr{2004}  \at{Predictive pore-scale modeling of two-phase flow in mixed wet media}.  \jt{Water Resources Research}  \bvol{40}~(7).

\bibitem[Vel{\'a}squez-Parra {\em et~al.\/}(2022)Vel{\'a}squez-Parra, Aquino, Willmann, M{\'e}heust, Le~Borgne \& Jim{\'e}nez-Mart{\'\i}nez]{velasquez2022sharp}
{\sc \au{Vel{\'a}squez-Parra, Andr{\'e}s}, \au{Aquino, Tom{\'a}s}, \au{Willmann, Matthias}, \au{M{\'e}heust, Yves}, \au{Le~Borgne, Tanguy} \& \au{Jim{\'e}nez-Mart{\'\i}nez, Joaqu{\'\i}n}} \yr{2022}  \at{Sharp transition to strongly anomalous transport in unsaturated porous media}.  \jt{Geophysical Research Letters}  \bvol{49}~(3),  \pg{e2021GL096280}.

\bibitem[Wang(2004)]{wang2004fundamental}
{\sc \au{Wang, Chao-Yang}} \yr{2004}  \at{Fundamental models for fuel cell engineering}.  \jt{Chemical Reviews}  \bvol{104}~(10),  \pg{4727--4766}.

\bibitem[Weller {\em et~al.\/}(1998)Weller, Tabor, Jasak \& Fureby]{weller1998tensorial}
{\sc \au{Weller, Henry~G}, \au{Tabor, Gavin}, \au{Jasak, Hrvoje} \& \au{Fureby, Christer}} \yr{1998}  \at{A tensorial approach to computational continuum mechanics using object-oriented techniques}.  \jt{Computers in Physics}  \bvol{12}~(6),  \pg{620--631}.

\end{thebibliography}

\end{document}